\documentclass[iop]{emulateapj}
\newcommand\degree{\degr}
\newcommand\degrees\degree

\DeclareSymbolFont{UPM}{U}{eur}{m}{n}
\DeclareMathSymbol{\umu}{0}{UPM}{"16}
\let\oldumu=\umu
\renewcommand\umu{\ifmmode\oldumu\else\math{\oldumu}\fi}
\newcommand\micro{\umu}
\renewcommand\micron{\micro m}
\newcommand\microns \micron

\let\oldsim=\sim
\renewcommand\sim{\ifmmode\oldsim\else\math{\oldsim}\fi}
\let\oldpm=\pm
\renewcommand\pm{\ifmmode\oldpm\else\math{\oldpm}\fi}
\newcommand\by{\ifmmode\times\else\math{\times}\fi}

\newbox{\wdbox}
\renewcommand\c{\setbox\wdbox=\hbox{,}\hspace{\wd\wdbox}}
\renewcommand\i{\setbox\wdbox=\hbox{i}\hspace{\wd\wdbox}}

\newcount\timect
\newcount\hourct
\newcount\minct
\newcommand\now{\timect=\time \divide\timect by 60
         \hourct=\timect \multiply\hourct by 60
         \minct=\time \advance\minct by -\hourct
         \number\timect:\ifnum \minct < 10 0\fi\number\minct}

\catcode`@=11

\newcommand\comment[1]{}
\newcommand\commenton{\catcode`\%=14}
\newcommand\commentoff{\catcode`\%=12}

\renewcommand\math[1]{$#1$}
\newcommand\mathshifton{\catcode`\$=3}
\newcommand\mathshiftoff{\catcode`\$=12}

\comment{the backslash is necessary}

\comment{alignment tab}

\let\atab=&
\newcommand\atabon{\catcode`\&=4}
\newcommand\ataboff{\catcode`\&=12}

\let\oldmsp=\sp
\let\oldmsb=\sb
\def\sp#1{\ifmmode
           \oldmsp{#1}%
         \else\strut\raise.85ex\hbox{\scriptsize #1}\fi}
\def\sb#1{\ifmmode
           \oldmsb{#1}%
         \else\strut\raise-.54ex\hbox{\scriptsize #1}\fi}
\newbox\@sp
\newbox\@sb
\def\sbp#1#2{\ifmmode%
           \oldmsb{#1}\oldmsp{#2}%
         \else
           \setbox\@sb=\hbox{\sb{#1}}%
           \setbox\@sp=\hbox{\sp{#2}}%
           \rlap{\copy\@sb}\copy\@sp
           \ifdim \wd\@sb >\wd\@sp
             \hskip -\wd\@sp \hskip \wd\@sb
           \fi
        \fi}
\def\msp#1{\ifmmode
           \oldmsp{#1}
         \else \math{\oldmsp{#1}}\fi}
\def\msb#1{\ifmmode
           \oldmsb{#1}
         \else \math{\oldmsb{#1}}\fi}
\def\supon{\catcode`\^=7}
\def\supoff{\catcode`\^=12}
\def\subon{\catcode`\_=8}
\def\suboff{\catcode`\_=12}
\def\supsubon{\supon \subon}
\def\supsuboff{\supoff \suboff}

\newcommand\actcharon{\catcode`\~=13}
\newcommand\actcharoff{\catcode`\~=12}

\newcommand\paramon{\catcode`\#=6}
\newcommand\paramoff{\catcode`\#=12}

\comment{And now to turn us totally on and off...}

\newcommand\reservedcharson{\commenton \mathshifton \atabon \supsubon \actcharon
	\paramon}

\newcommand\reservedcharsoff{\commentoff \mathshiftoff \ataboff
	\supsuboff \actcharoff \paramoff}

\catcode`@=12
\reservedcharsoff

\reservedcharson

\comment{ Must have ONLY ONE of these... trust these macros, they work

}

\newcommand{\squishlist}{
 \begin{list}{$\bullet$}
  { \setlength{\itemsep}{1pt}
     \setlength{\parsep}{0pt}
     \setlength{\topsep}{3pt}
     \setlength{\partopsep}{0pt}
     \setlength{\leftmargin}{2.0em}
     \setlength{\labelwidth}{1.5em}
     \setlength{\labelsep}{0.5em} } }

\newcommand{\squishend}{
  \end{list}  }

\reservedcharson
\actcharon

\usepackage{natbib}
\usepackage{amsmath}
\usepackage{xcolor}

\newcommand{\Kepler}{{\sl Kepler}\ }

\newcommand{\be}{\begin{equation}}
\newcommand{\ee}{\end{equation}}
\newcommand{\Q}{\mathcal{Q}}
\newcommand{\C}{\mathcal{C}}

\newcommand{\review}[1]{\textcolor{black}{#1}}

\slugcomment{}
\shorttitle{ Information theory for exoplanets }
\shortauthors{Gilbert \& Fabrycky}

\begin{document}
\title{
An information theoretic framework for classifying exoplanetary system architectures
}
\author{
Gregory J. Gilbert\altaffilmark{1} \&
Daniel C. Fabrycky\altaffilmark{1}
}
\altaffiltext{1}{Department of Astronomy and Astrophysics, University of Chicago, 5640 S. Ellis Ave., Chicago, IL 60637, USA}
\email{Email: gjgilbert@uchicago.edu}

\begin{abstract}

We propose several descriptive measures to characterize the arrangements of planetary masses, periods, and mutual inclinations within exoplanetary systems. These measures are based in complexity theory and capture the global, system-level trends of each architecture. Our approach considers all planets in a system simultaneously, facilitating both intra-system and inter-system analysis. We find that based on these measures, \textit{Kepler's} high-multiplicity ($N\geq3$) systems can be explained if most systems belong to a single intrinsic population, with a subset of high-multiplicity systems ($\sim20\%$) hosting additional, undetected planets intermediate in period between the known planets. We confirm prior findings that planets within a system tend to be roughly the same size and approximately coplanar. We find that forward modeling has not yet reproduced the high degree of spacing similarity (in log-period) actually seen in the \Kepler data. Although our classification scheme was developed using compact \Kepler multis as a test sample, our methods can be immediately applied to any other population of exoplanetary systems. We apply this classification scheme to (1) quantify the similarity between systems, (2) resolve observational biases from physical trends, and (3) identify which systems to search for additional planets and where to look for these planets.

\end{abstract}


\section{Introduction}\label{sec:Intro}

Describing exoplanetary system architectures is an inherently complex task. Although the basic observables for each individual planet (mass, radius, period, inclination, and eccentricity) are relatively straightforward to measure and compare, there is no single preferred way to combine these variables in order to assess system-level trends. Compounding the problem, the number of possible combinations of parameters scales rapidly with the number of planets in a system. Furthermore it is not always clear how to compare, say, a 2-planet system to a 6-planet system.

The standard approach has been to consider pairwise statistics, typically of adjacent planet pairs. For example, one might compute the ratios of planet masses, radii, orbital periods, or velocity-scaled transit durations. This approach has been quite fruitful, leading to powerful constraints on the distributions of planet sizes and orbital arrangements in multiplanet systems \citep{Lissauer2011, Ciardi2013, Fabrycky2014, WinnFabrycky2015, Millholland2017, Wang2017, Weiss2018}. However, this pairwise approach has two major drawbacks. First, this approach implicitly assumes that a system's global architecture is well described by pairwise statistics, which may or may not be the case. In reality, complex systems often display emergent properties that arise from interactions between independent components. Second, employing pairwise ratios does little to reduce the dimensionality of the problem at hand, and if non-adjacent planet pairs are considered actually \textit{increases} the dimensionality of the problem.

A better approach would be to consider all planets in a system simultaneously using higher-order statistics that reduce the number of dimensions to a tractable level. Consider, for example, planet masses. Rather than tallying $N$ masses and $N(N-1)/2$ mass ratios, we aim to summarize the partitioning of mass between planets as a single number. For a 5-planet system, our approach would reduce the dimensionality by four, whereas the standard approach would increase the dimensionality by ten if all masses and mass ratios were considered independently. The benefits will be at least as substantial for any other quantity considered.

Our goal then is this: we aim to identify a small number of parameters which parsimoniously capture the global architecture of an exoplanetary system.

Beyond mathematical convenience, there is good physical motivation for prioritizing the system over the individual planets. Because all planets in a system share a common formation history, their properties are intrinsically linked, and thus planets do not constitute independent samples. A corollary is that any population level studies that treat planets independently will be inherently biased by correlations between the input variables. Stars, however - and by extension systems - do constitute independent samples, at least to first order when ignoring the effects of cluster environment or stellar multiplicity on planet formation. We therefore argue for a subtle yet radical shift in perspective: rather than treating the \textit{planet} as the fundamental unit of exoplanet science, we treat the \textit{system} as the fundamental unit.

But how do we reduce a complex system of planets to just a few numbers? Fortunately, there is a branch of mathematics well-suited to the task, aptly named ``complexity theory,'' which is itself an extension of information theory. In this context, the word ``complex'' refers to any system which has distinct properties that arise from the interactions between components or patterns which are not regarded as simple \citep{LMC2010}. Under this definition, planetary systems are indeed complex, and so we believe that complexity theory is the right tool for the job. 

This paper is organized as follows. In \S\ref{sec:InfoTheory} we review some relevant ideas from the information theory literature. In \S\ref{sec:Measures} we propose several new measures for classifying exoplanetary systems. In \S\ref{sec:Clustering} we combine these measures and search for clusters of distinct system type. In \S\ref{sec:Synthetic} we compare real \Kepler systems to synthetic populations. In section \S\ref{sec:Trends} we investigate whether the observed trends can be explained a subpopulation of systems with as-yet undetected planets.  In \S\ref{sec:Summary} we summarize our results and discuss possible modifications to our classification scheme.

\section{Overview of information theory}\label{sec:InfoTheory}

Because the application of information theory is relatively novel to the field of astrophysics in general, and to the study of exoplanets in particular, here we review some foundational ideas from the information theory literature. The majority of this review is borrowed from \citet*{LMC2010}, and so we refer the interested reader there for further detail.

\subsection{The big idea: convex complexity}

The seminal work of information theory came from Claude \citet{Shannon1949}, who linked the information content of a system to its entropy via the now-famous ``Shannon information,'' or ``Shannon entropy,'' defined as

\begin{equation}\label{H}
    H \equiv -\sum_{i=1}^N p_i\log p_i
\end{equation}

where $H$ is the Shannon information and $p_i$ are occupancy probabilities for the $N$ possible states of the system\footnote{Here the word ``system'' is used in the general sense of any physical system, and not in the particular sense of a planetary system.} being studied. The units of entropy are variously taken to be bits (for $\log_2$), nats ($\log_e$) or bans ($\log_{10}$). Although system behavior is identical regardless of units, there is no single standard for which base to use, and so one should take care when comparing results, especially when doing so across scientific disciplines. Unless otherwise noted, we use natural logarithms throughout this work.

In order to gain an intuition for how Shannon information behaves, it is instructive to examine two end-member cases: an ideal gas (maximum entropy) and a perfect crystal (zero entropy). For an ideal gas, each energy state has equal probability, and so Equation \ref{H} is maximized, whereas for a perfect crystal a single energy state has $p_i \rightarrow 1$, and so Equation \ref{H} is minimized. From the perspective of information theory, one can interpret this result as follows. The information content required to describe a perfect crystal is minimal - one needs only a single number (or perhaps a small handful of numbers) to specify the lattice bond length(s) in order to capture the entire structure of the crystal. For an ideal gas, on the other hand, one must specify the position and momentum of every individual particle in order to describe the full structure. The solution from statistical mechanics, describing the gas as a distribution over momentum states, is exactly what its name implies - statistical - and does not truly capture the full information content of the physical system.

The mind may rebel at the notion that an ideal gas holds maximal information content. ``Obviously'' the entropy of the system is at a maximum, yet at the same time it is ``obvious'' that little useful information can be retrieved from the individual particles. For a fully randomized system, even though the formal information content (i.e. the entropy) has been maximized, the extractable information has been minimized. Clearly, entropy alone is not sufficient to capture the behavior of these systems.

A complementary statistic to entropy is the \textit{disequilibrium}, defined as

\begin{equation}\label{D}
    D \equiv \sum_{i=1}^N (p_i - \frac{1}{N})^2
\end{equation}

For our end-member cases, $D$ is minimized for the ideal gas, in which every energy state has equal probability, and maximized for the perfect crystal, in which a single energy states dominates completely. Thus, disequilibrium shows qualitatively inverse behavior to entropy.

Our intuition tells us that both an ideal gas (fully randomized) and a perfect crystal (fully ordered) have zero complexity. But zero complexity is \textbf{not} synonymous with zero entropy; a zero entropy system will have zero complexity, but zero complexity system need not have zero entropy. Because entropy is often colloquially described as “chaos” or “disorder”, the distinction between complexity and entropy can sometimes become blurred. So, at the risk of redundancy, we reiterate that our aim is to distinguish between systems with low versus high complexity, which may not always be the same as distinguishing between systems with low versus high entropy.

\begin{figure}
    \centering
    \includegraphics[width=0.45\textwidth]{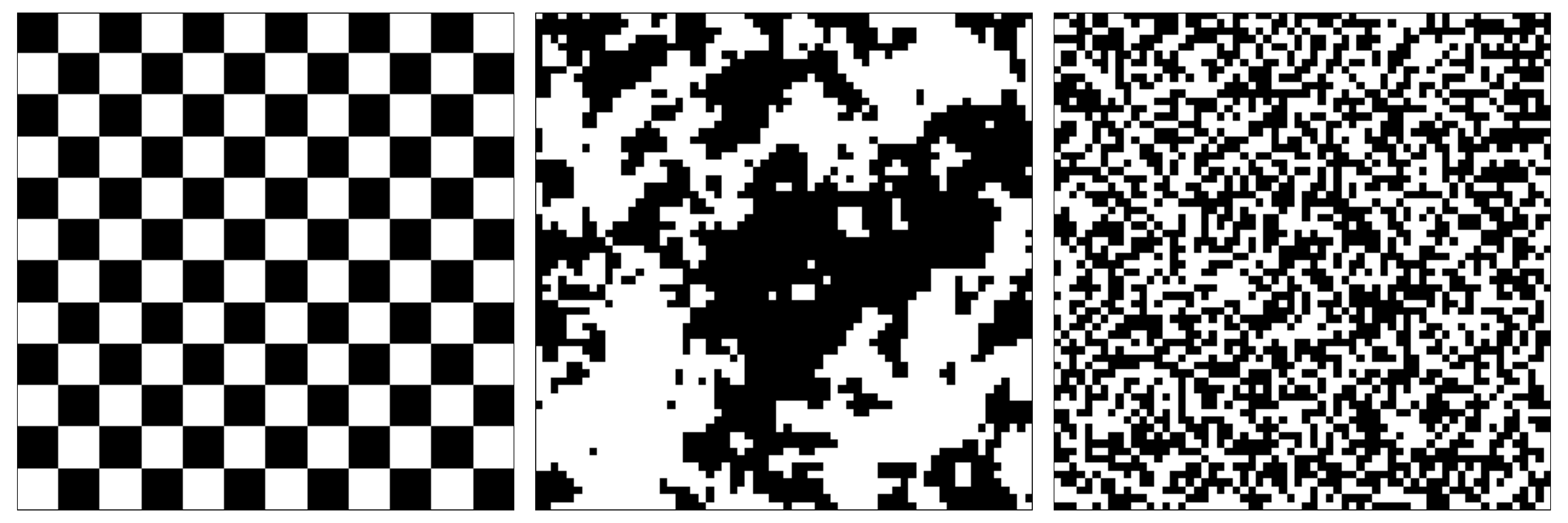}
    \caption{An illustration of the intuitive notion of complexity. Both complete order (left panel) and total randomness (right panel) have low complexity, while the intermediate state (middle panel) displays patterns that we might intuitively call complex.}
    \label{fig:complexity_examples}
\end{figure}

What we desire then is a measure of complexity which goes to zero for both perfectly ordered and perfectly random systems, peaking at some maximum complexity for an intermediate state where a balance between entropy and equilibrium is achieved. An example of such a situation is shown in Figure \ref{fig:complexity_examples}. Using information theory parlance, measures which meet the above criteria are sometimes called \textit{convex} complexity measures. Our discussion of information theory has so far implicitly treated convex complexity measures as if these are the \textit{only} way to define complexity. In fact, some definitions of complexity are defined as monotonic functions of entropy, in direct contradiction to our arguments above. However, these definitions tend to arise primarily in the field of computer science \citep[cf.][]{Chaitin1966, Kolmogorov1968}. In the social, biological, and physical sciences, convex complexity measures dominate, and so as not to confuse matters, we will restrict ourselves to consideration of convex measures.

\begin{figure}
    \centering
    \includegraphics[width=0.45\textwidth]{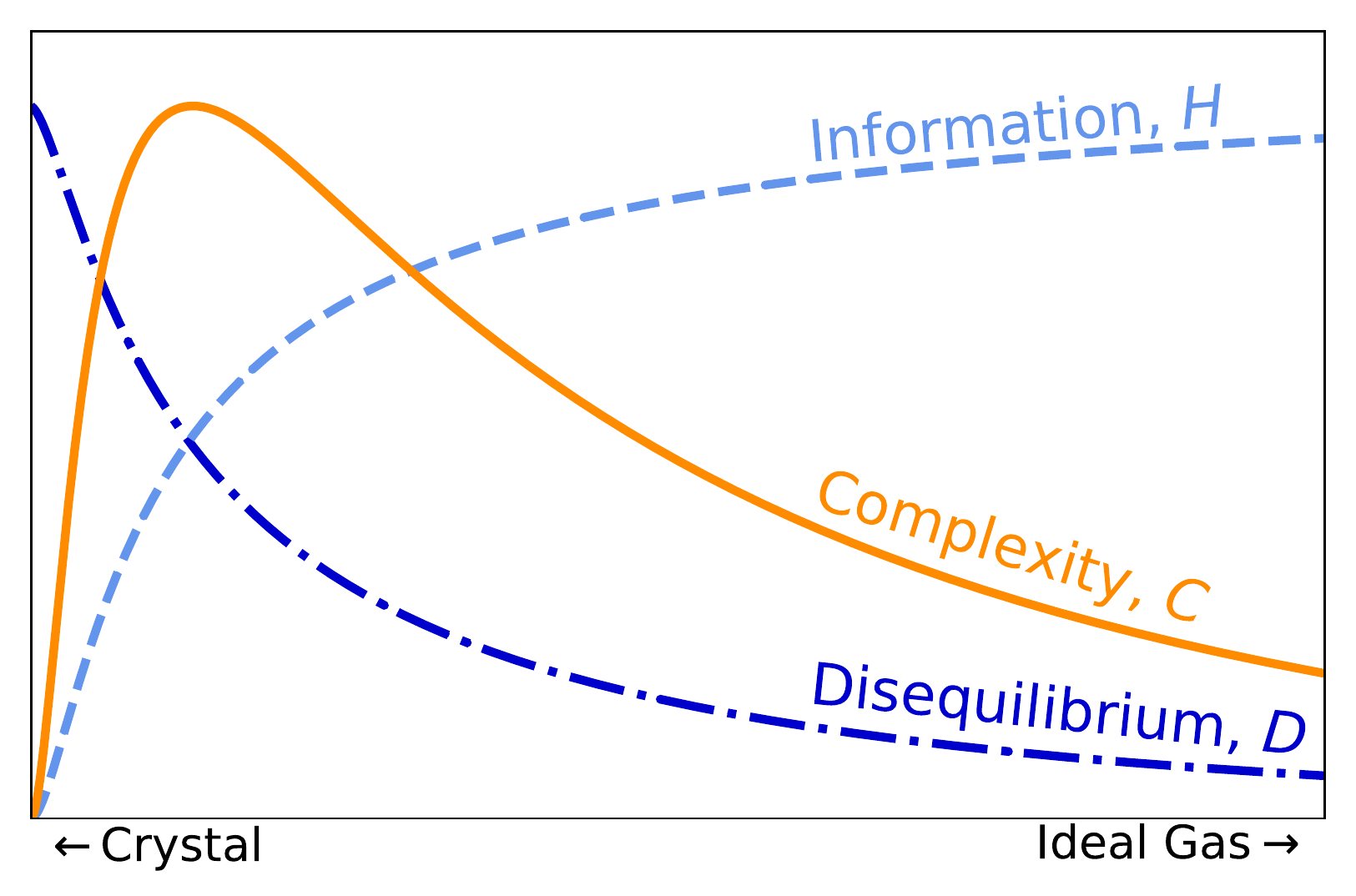}
    \caption{A schematic representation of the behavior of information, $H$, disequilibrium, $D$, and LMC complexity, $C = H \cdot D$. Complexity reaches a maximum at a midpoint between complete order (perfect crystal) and total randomness (ideal gas). Any measure of convex complexity must satisfy these boundary conditions. Adapted from \citet*{LMC2010}.}
    \label{fig:CHD_curve}
\end{figure}

There are two convex complexity measures commonly used in the physical sciences. The first was introduced by \citet*{LMC1995}, who defined complexity as the product of entropy and disequilibrium:

\begin{equation}\label{C}
    C \equiv H \cdot D = -K \left(\sum_{i=1}^N p_i\log p_i \right)
    \cdot \left( \sum_{i=1}^N (p_i - \frac{1}{N})^2 \right)
\end{equation}

where $K$ is a positive, real constant. $C$ is a statistical measure of complexity and is sometimes called LMC complexity (after the authors; no relation to the Large Magellanic Cloud). A schematic representation of the behavior of $H$, $D$, and $C$ is shown in Figure \ref{fig:CHD_curve}.

\begin{figure}
    \centering
    \includegraphics[width=0.45\textwidth]{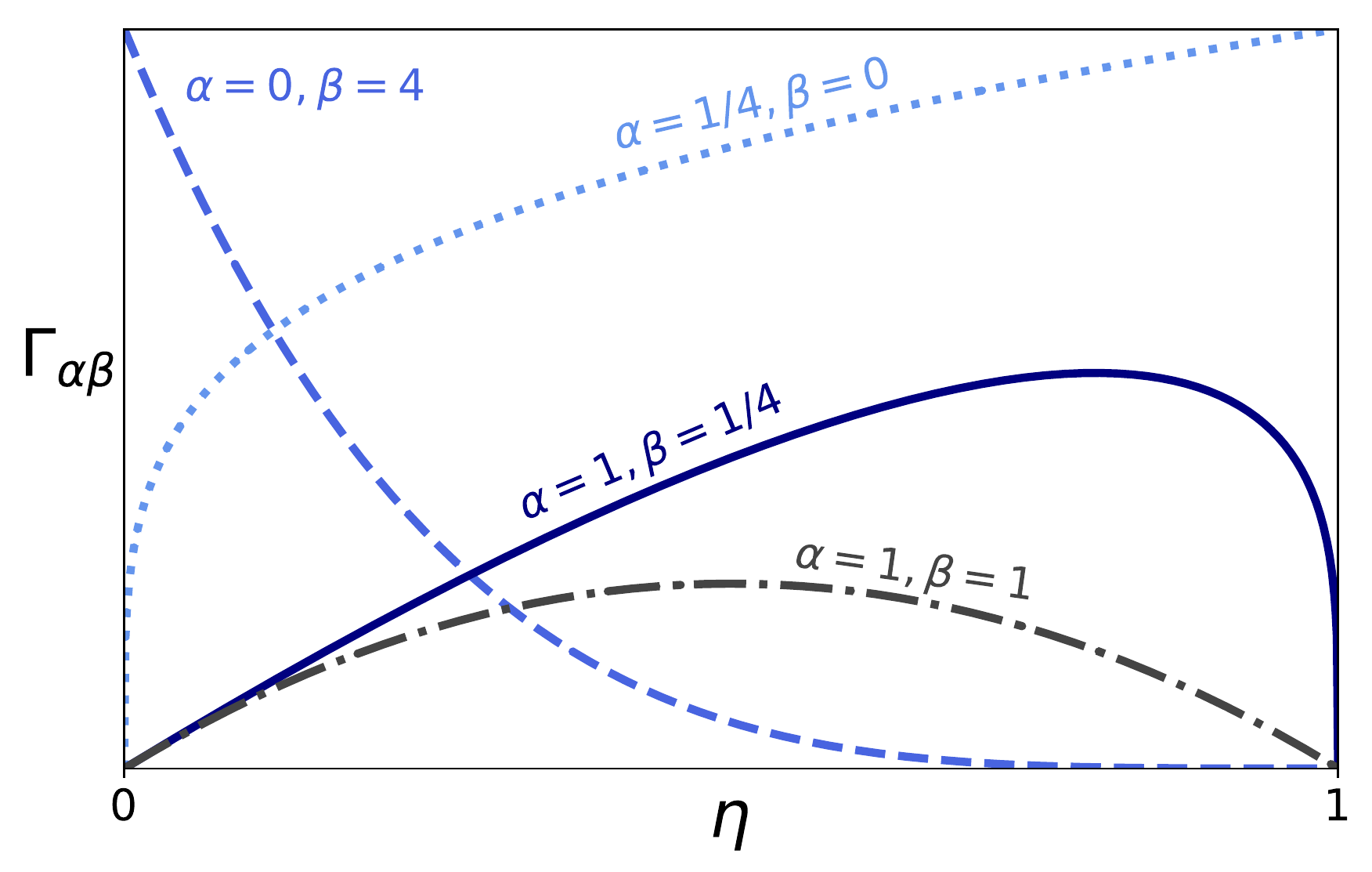}
    \caption{SDL complexity, $\Gamma$, vs disorder, $\eta$. Setting either $\alpha$ or $\beta$ equal to zero results in a monotonic function $\Gamma(\eta)$, whereas setting both $\alpha$ and $\beta$ to positive, nonvanishing values produces a convex complexity curve with varying degrees of skewness. See Equations \ref{eq:Gamma} \& \ref{eq:disorder} for definitions of quantities. Adapted from \citet*{Shiner1999}. }
    \label{fig:SDL_curve}
\end{figure}

The second measure, a quantity closely related LMC's \textit{statistical} measure of complexity, is the \textit{simple} measure of complexity proposed by \citet*{Shiner1999}. The so-called SDL complexity, $\Gamma$, is defined as

\begin{equation}\label{eq:Gamma}
    \Gamma \equiv \eta^\alpha(1-\eta)^\beta
\end{equation}

\begin{equation}\label{eq:disorder}
    \eta = H/H_{max}
\end{equation}

where $\alpha$ and $\beta$ parameterize the relative weighting of disorder, $\eta$, vs order, $1-\eta$, and $H_{max}$ is the maximum entropy achievable by the system being studied. Figure \ref{fig:SDL_curve} shows how $\Gamma$ behaves for a few values of $\alpha$ and $\beta$. For positive, nonvanishing values of $\alpha$ and $\beta$, SDL complexity has the qualitative feature of going to zero near perfect order and near total randomness, peaking at a maximum somewhere in the middle. Note, however, that certain choices of $\alpha$ or $\beta$ can also produce a complexity curve in which $\Gamma$ is a monotonic function of entropy (i.e. not convex). For an in-depth comparison of $C$ and $\Gamma$, see \citet{Panos2006}, who explore the behavior of these two measures as they relate to atomic structure. 

When applying the idea of convex complexity to exoplanetary system architectures, we opt to use $C$ over $\Gamma$ for two reasons. First, $\Gamma$ derives directly from the entropy, without explicit consideration of disequilibrium. Because a dominant feature of exoplanet architectures appears to be their frequent nearness to equipartitioning, we believe that $C$ is a more appropriate choice for characterizing complexity. Second, whereas calculating $C$ - to within a normalization factor - provides a fixed value dependent only on the number and occupancy probabilities of allowed states, calculating $\Gamma$ includes two additional parameters, $\alpha$ and $\beta$, which allows for a higher level of modeling flexibility that we do not believe is warranted by the current quality of available data. However, we mention $\Gamma$ here for completeness and because others might find use of it in the future.

As an aside, it is worth noting that all of our above discussions have been limited to a broad class of ideas that fall under the umbrella of \textit{algorithmic} complexity. These measures all derive in one form or another from Shannon entropy. Roughly speaking, there are two other main branches of complexity theory: \textit{deterministic} complexity, commonly known as chaos theory, and \textit{aggregate} complexity, which focuses on how individual elements create complex patterns and systems, such as those found in the transmission of epidemic diseases or in the migration patterns of birds. The distinctions between these three categories are somewhat arbitrary and sometimes ill-defined, and each branch shares many overlapping ideas with the other two. Nevertheless, such distinctions can be useful to make when diving into the vast literature of information theory. We direct the reader to three excellent reviews by \citet{May1976}, \citet{Manson2001}, and \citet{Lansing2003} for further exploration of these ideas.

\subsection{Application to astrophysics}

Before defining our new measures to describe exoplanetary system architectures, we outline our philosophical approach to this problem.

Our technique is descriptive and not tied to any underlying theories of planet formation or orbital dynamics. Why? Because we wish to characterize patterns in the data without introducing biases that might arise from making physical assumptions. Our philosophy is ``describe first, explain later.''

\begin{figure}
    \centering
    \includegraphics[width=0.45\textwidth]{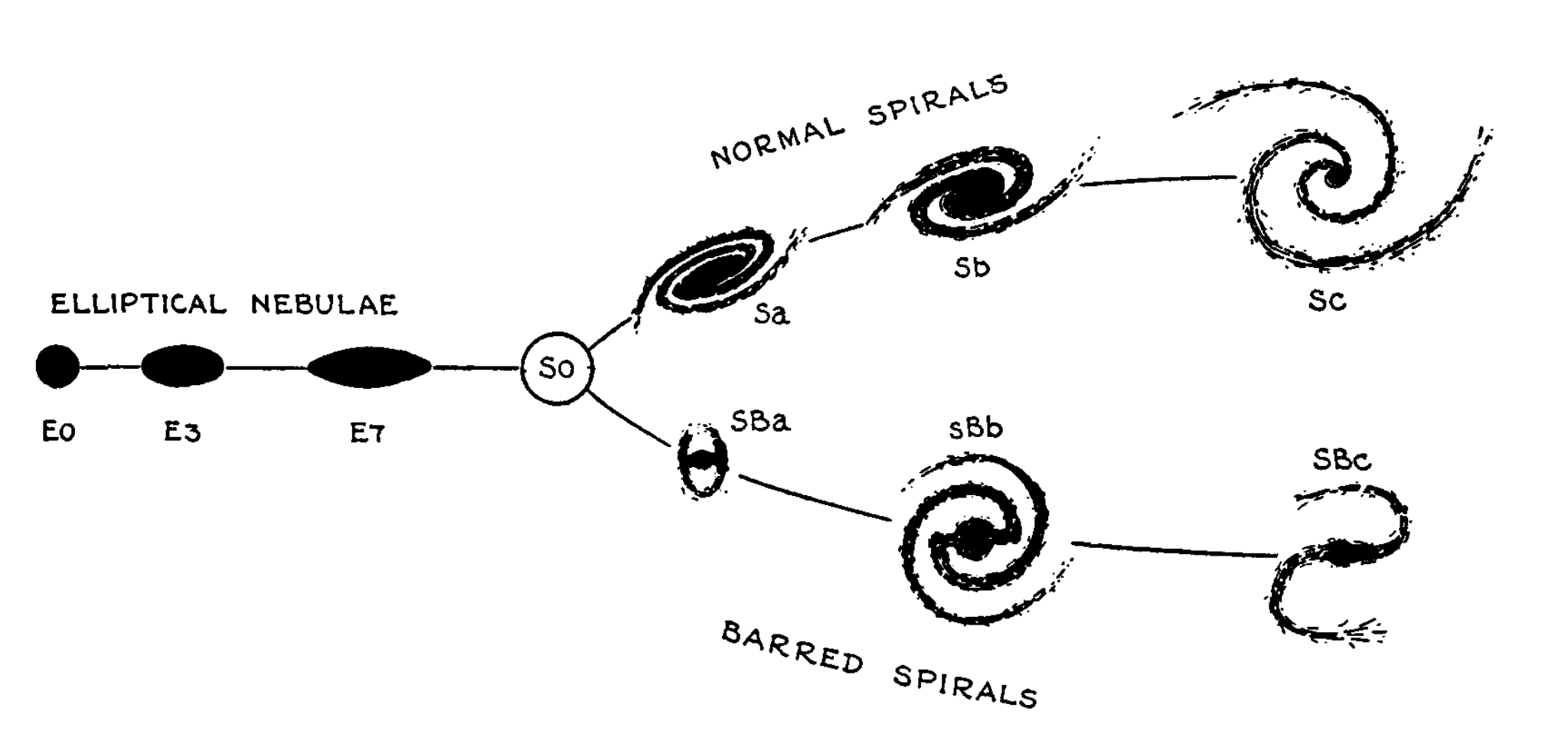}
    \caption{Hubble's ``tuning fork'' classification of galaxies. Even though Jeans' contemporaneous explanation of this classification did not withstand the test of time, Hubble's morphological scheme nonetheless captured some of the most important features of galaxy evolution and effectively distinguished galaxies into physically meaningful groups. This classification was purely descriptive and even grouped galaxies by an intuitive notion of complexity. We seek to take a similar approach to classifying exoplanet systems. Figure reproduced from \citet{Hubble1936}.}
    \label{fig:HubbleSequence}
\end{figure}

As a motivating example, consider Edwin Hubble's development of a morphological classification scheme for galaxies \citep[Figure \ref{fig:HubbleSequence};][]{Hubble1926, Hubble1936}. To quote Hubble directly, while developing this scheme, ``deliberate effort was made to find a descriptive classification which should be entirely independent of theoretical considerations. The results are almost identical with the path of development derived by Jeans from purely theoretical investigations...However, the basis of the classification is descriptive and entirely independent of any theory." Even though Jeans' explanation did not withstand the test of time, Hubble's sequence nonetheless captured many of the most important features of galaxy structure. Indeed, Hubble's purely morphological description was so successful that it remains relevant nearly a century later!

Most modern efforts to classify galaxies are interpreted in relation to Hubble's original scheme. Although some progress has recently been made towards automatic classification of galaxies based either on spectra \citep{SanchesAlmeida2010} or morphology \citep{Shamir2009}, classification ``by-eye'' remains one of the most reliable methods for classification, aided by the large volume of data available from surveys such as the Sloan Digital Sky Survey \citep{York2000} and crowd-sourced analysis efforts such as Galaxy Zoo \citep{GalaxyZoo1, GalaxyZoo2}. These various classification methods all rely on a small handful of observable feature (the presence or lack of spiral arms and bars, bulge-to-disk ratios, luminosity, emission spectra, etc.) and assign physical interpretation of these features (e.g. rate of star formation) after - not before - classification. The field of exoplanets needs a framework like this so that we can first identify the critical features of systems and then assign physical meaning to those features.

We adopt a similar approach here, eschewing the theoretical framework of planet formation in favor of a purely descriptive technique. Like Hubble, we are aware of the existing theoretical explanations for observed structure (i.e. planet formation models), but we make deliberate effort to develop our classification scheme to be independent of this or any other theory. If our methods are sound, they should naturally capture the important features of the population of systems. In order to be accessible to the community, we strive to make our measures straightforward to calculate and easy to interpret.

\section{Description of our classification scheme}\label{sec:Measures}
Here we propose several measures to quantify the global structure of planetary systems. Some are modifications of quantities which already exist in the literature, while others are new as of this work. Our proposed measures are as follows:

\begin{enumerate}
    \item \textit{Dynamical mass}, $\mu$, sets the overall mass scale of the system;
    \item \textit{Mass partitioning}, $\Q$, captures the variance in masses between planets;
    \item \textit{Monotonicity}, $\mathcal{M}$, describes the size ordering of the planets;
    \item \textit{Characteristic spacing}, $\mathcal{S}$, is the average separation between planets in mutual Hill radii;
    \item \textit{Gap complexity}, $\C$, summarizes the relationships between orbital periods;
    \item \textit{Flatness}, $f$, is related to the scatter in mutual inclinations;
    \item \textit{Multiplicity}, $N$, is the observed number of planets in a system.
\end{enumerate}

\begin{figure*}
    \centering
    \includegraphics[width=0.90\textwidth]{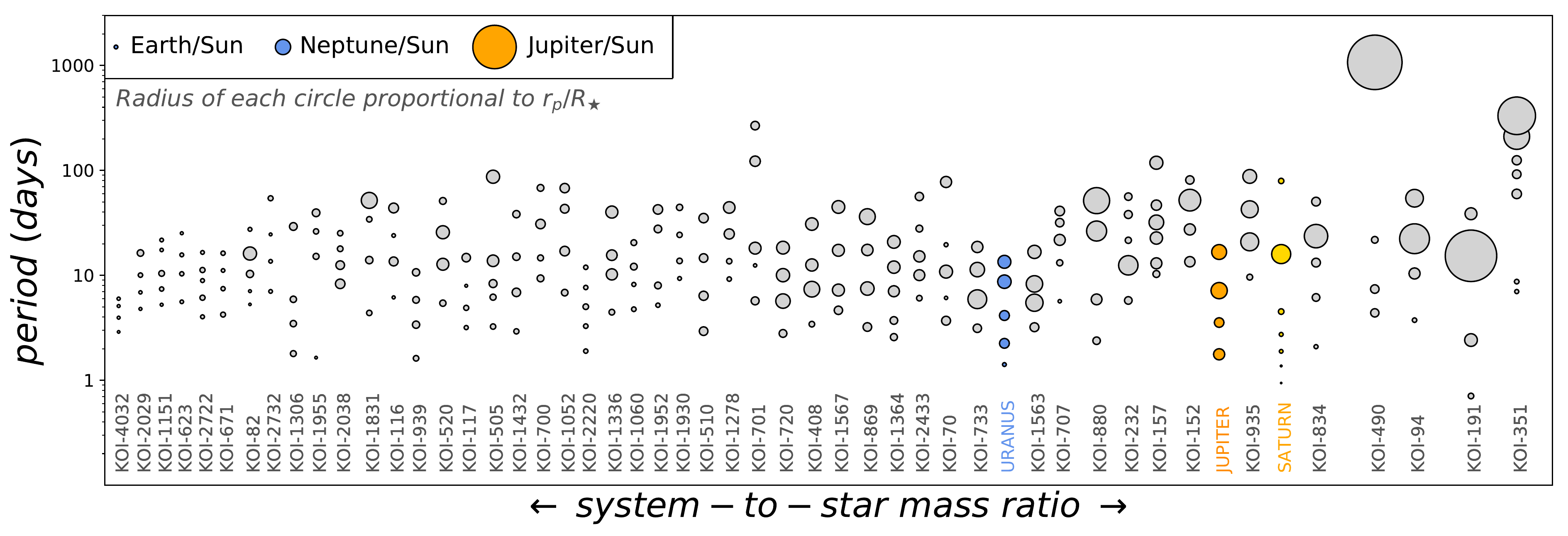}
    \caption{Gallery of 4+ planet systems considered in this study. Systems are arranged from left-to-right by increasing dynamical mass. The giant moon systems of Jupiter, Saturn, and Uranus are shown for comparison.}
    \label{fig:gallery}
\end{figure*}

When evaluating these measures on real systems, we use the catalogue from the California Kepler Survey \citep[CKS;][]{Johnson2017, Petigura2017}. To ensure a high-quality sample, we cross-match all candidate planets with \Kepler Data Release 25 \citep[DR25;][]{Thompson2018}. We next make a few reasonable cuts to remove false positives, grazing transits ($b>1-r_p/R_{\star}$) and low signal-to-noise ($SNR<7.1$) objects. Finally, in order to ensure that all stellar characterization is accurate, we apply restrictions on stellar radius, temperature, ``isochrone parallax,'' and dilution following the procedures described in section 4.2 of \citet{Fulton2018} and using the stellar companion catalogue of \citet{Furlan2017}. After applying these cuts, we are left with 864 planets in 335 multiplanet systems. Of these, 452 planets are found in 129 high-multiplicity ($N \geq 3$) systems. A gallery of 4+ planet systems considered in this study is shown in Figure \ref{fig:gallery}, and a summary of system level statistics is given in Table \ref{tab:SystemSummary}. All of the stars in the sample have been well characterized by Gaia \citep{Gaia2016} and CKS spectroscopy, which provide tight constraints on stellar density, thus allowing us to more precisely determine transit durations and, by extension, mutual inclinations. Furthermore, the high precision on stellar parameters translates into correspondingly high precision on planetary periods and radii.

\begin{table*}[tb]
\centering
\caption{Summary of system-level measures: dynamical mass $\mu$, mass partitioning $\Q$, monotonicity $\mathcal{M}$, characteristic spacing $\mathcal{S}$, gap complexity $\C$, flatness $f$, and multiplicity $N$. Each system is identified by its Kepler Object of Interest (KOI) number. The final column, labeled ``pop.'' identifies which subpopulation each system belongs to, as described in sections \ref{sec:Clustering} \& \ref{sec:Trends}. Some variables are undefined for systems with $N < 3$, but systems with $N=2$ are included for completeness. The full version of this table will be available in machine-readable form.}
    \begin{tabular}{c|c|c|c|c|c|c|c|c}\label{tab:SystemSummary}
    KOI & $\log_{10}\mu$ & $\Q$ & $\mathcal{M}$ & $\mathcal{S}$ & $\C$ & $f$ & $N$ & pop. \\
    \hline
    K00041 & -4.510 & 0.063 & -0.20 & 26.9 & 0.181 & 0.148 & 3 & 0 \\
    K00046 & -3.907 & 0.852 & -0.92 & 10.4 & - & 0.051 & 2 & - \\
    K00070 & -4.037 & 0.150 & -0.07 & 22.6 & 0.244 & 0.163 & 5 & 0 \\
    K00072 & -4.449 & 0.133 & 0.36 & 76.3 & - & 0.163 & 2 & - \\
    K00082 & -4.386 & 0.452 & 0.60 & 18.4 & 0.054 & 0.062 & 5 & 0 \\
    K00085 & -4.348 & 0.059 & 0.20 & 19.5 & 0.677 & 0.007 & 3 & 1 \\
    K00094 & -3.396 & 0.312 & 0.60 & 15.9 & 0.025 & 0.074 & 4 & 0 \\
    \vdots & \vdots & \vdots & \vdots & \vdots & \vdots & \vdots & \vdots & \vdots \\
    \end{tabular}
\end{table*}

Because masses are not available for the majority of \Kepler planets, we convert radii to mass using the probabilistic mass-radius-period relations of \citet{Neil2019}. These relations are similar to the probabilistic forecasting of \citet{Chen2017} but also incorporate information on orbital period and stellar insolation. Although there is considerable scatter in the mass-radius-period relation, with a sample of several hundred planets we expect this scatter to marginalize out so that we can still see statistical trends within the population of systems. We choose to work with mass rather than directly working with radius because mass is the more fundamental quantity tied to planet formation. In order to maintain a homogeneous sample, we do not use masses derived directly from radial velocities or transit timing variation measurements, even when such masses are available. These mass measurements are, however, incorporated into the probabilistic model of \citet{Neil2019}.

Our decision to use a sample of only transiting planets is driven by a desire for convenient comparison with previous works, in particular \citet{Weiss2018}, as well as by the large number of systems and sensitivity to individual planet inclinations. However, our techniques are not tied to any specific detection method, and so the measures described in this section could be easily generalized to a population of planets detected by any other method or even some heterogeneous combination of detection methods. To keep things simple, we stick to a homogeneous CKS catalogue and save the application of these measures to other exoplanet populations for future work.

In the subsections which follow we describe each of our proposed system level measures in turn.

\subsection{Dynamical mass, $\mu$}

Mass is arguably the most fundamental property of an individual planet, and so describing the mass scale of each system a natural place to start when characterizing exoplanetary system architectures. The \textbf{dynamical mass} of the system is defined as 

\begin{equation}
    \mu \equiv \sum_{i=1}^N m_i/M_{\star}
\end{equation}

where $m_i$ are planet masses and $M_{\star}$ is the stellar mass, with the term ``dynamical mass'' taken following \citet{JontofHutter2016}. We choose to report mass as the system-to-star mass ratio rather than the simpler total integrated mass because planet formation and orbital dynamics are more closely related to disk-to-star and planet-to-star mass ratios than to total mass. Indeed, we note that the distribution of dynamical masses is conspicuously peaked near $\mu \approx 10^{-4}$ (Figure \ref{fig:masspartitioning}), commensurate with the common dynamical masses of the Jovian, Saturnian, and Uranian moon systems \citep[e.g.][]{Mosqueira2003, Canup2006}, hinting at a common formation pathway for exoplanet systems and giant planet satellites \citep{Chiang2013, Miguel2019}.

\begin{figure}
    \centering
    \includegraphics[width=0.45\textwidth]{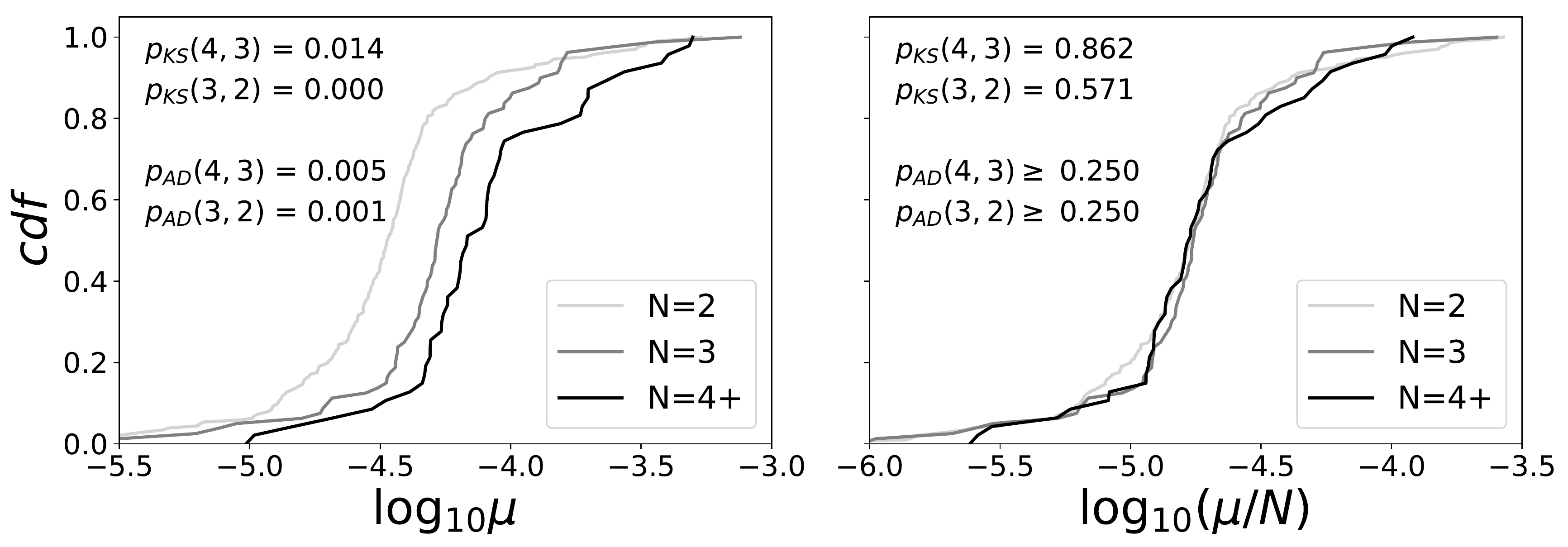}
    \caption{Cumulative density function of $\log_{10}\mu$ (left panel) and $\log_{10}(\mu/N)$ (right panel). Both Kolmogorov-Smirnov and Anderson-Darling tests indicate that while $\log\mu$ is drawn from different distributions for different multiplicities, when $\mu$ is normalized by multiplicity all systems appear to have an average planet size drawn from the same underlying distribution.}
    \label{fig:mu_cdf}
\end{figure}

The cumulative density function (cdf) of $\mu$ and $\mu/N$ for 2, 3, and 4+ planet systems are shown in Figure \ref{fig:mu_cdf}. We find that both the Kolomogorov-Smirnov test statistic (K-S test) and Anderson-Darling test statistic (A-D test) indicate that while $\log\mu$ is drawn from different distributions for different multiplicities, when $\mu$ is normalized by multiplicity all systems appear to be drawn from the same underlying distribution. The straightforward interpretation is that the average planet size (relative to host star) is the same for all multiplicities $N\geq2$. We further hypothesize that these dynamical mass variations between multiplicities indicate that many of the lower multiplicity systems host additional undetected planets.

\subsection{Mass partitioning, $\Q$}

We define the \textbf{mass partitioning} of a system as

\begin{equation}\label{Q}
    \Q \equiv \left(\frac{N}{N-1}\right) \cdot \left( \sum_{i=1}^N (m_i^* - \frac{1}{N})^2 \right)
\end{equation}

\begin{equation}\label{eq:m*}
    m_i^* = m_i / \sum_i^N m_i
\end{equation}

The second bracketed term in Equation \ref{Q} is simply the disequilibrium, $D$, calculated from Equation \ref{D} with the substitution $p_i \rightarrow m_i^*$, where $m_i^*$ is the normalized planet mass defined following Equation \ref{eq:m*}. In the language of occupancy probabilities, $m_i^*$ can be thought of as the probability that an infinitesimal mass element $dm$ resides in a particular planet. In simple terms, $m_i^*$ is the fraction of total system mass (excluding the star) contained in an individual planet. The prefactor $N/(N-1)$ normalizes $\Q$ to the range (0,1). Thus, any system with all equal mass planets will have $\Q=0$ while a system with one dominant giant planet and $N-1$ tiny planets will have $\Q\rightarrow 1$.

We note that our definition of mass partitioning is closely related to the intra-system mass dispersion measures of \citet{Millholland2017} and \citet{Wang2017}. Our present work was largely inspired by these two studies, and so we are indebted to them. \review{Indeed, many of the distance metrics defined in these papers account for full-system architecture by considering all adjacent pair ratios simultaneously.} The critical advantage of our method over these antecedent works is that our measure is more intuitive to interpret and is more explicitly linked to the global architecture of each system, which facilitates not only intra-system comparison by also inter-system comparison.

\begin{figure}
    \centering
    \includegraphics[width=0.45\textwidth]{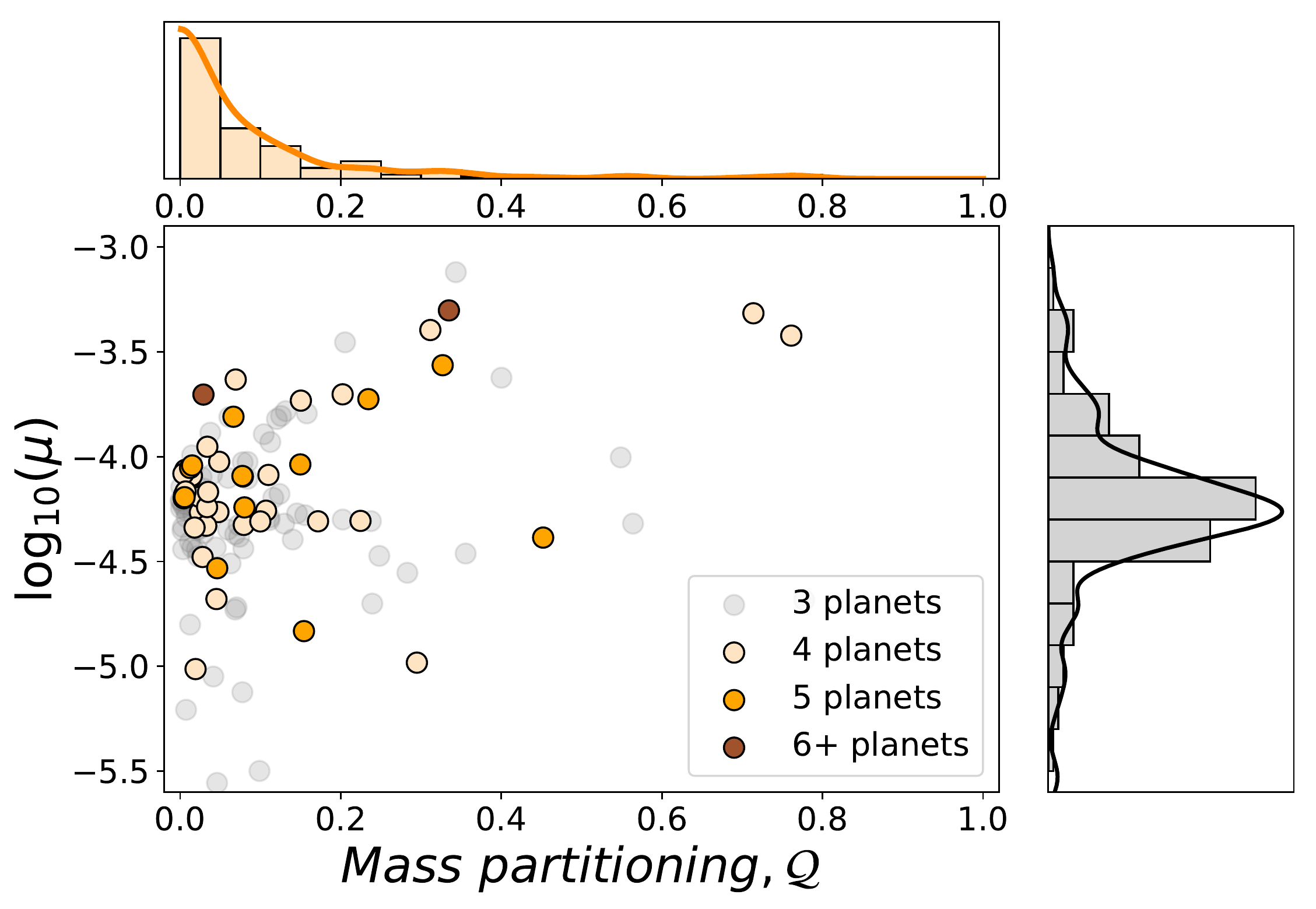}
    \caption{Distribution of mass partitioning, $\Q$, vs. dynamical mass $\mu$. Each 1-dimensional histogram is overplotted with a Gaussian kernel density estimate (KDE), the bandwidth for which was selected using Silverman's rule \citep{Silverman1986}. We find that the majority of systems have low $\Q$, indicating that planets within a system tend to be the same size. Among the few systems with large $\Q$, many also have high values of $\mu$, which is to be expected based on the definition of $\Q$. Conversely, we find few planets with large $\Q$ but low $\mu$, indicating that if giant planets do form, runaway gas accretion occurs unequally between planets in a system.}
    \label{fig:masspartitioning}
\end{figure}

The most striking feature of the distribution of $\Q$ (Figure \ref{fig:masspartitioning}) is that nearly all systems show a high degree of uniformity of planet sizes, i.e. low $\Q$, confirming previous results \citep{Millholland2017, Wang2017, Weiss2018}. There is no strong correlation between dynamical mass, $\mu$ and mass partitioning, $\Q$, although for relatively massive systems ($\log_{10}\mu \gtrsim -3.8$) there is a positive correlation between $\Q$ and $\log_{10}\mu$, a result which is expected under the runaway gas accretion model of giant planet formation \citep{Zhou2005}.

We find no obvious difference in the $\Q$ distribution of 3 planet vs 4+ planet systems as based on the results of K-S and A-D tests, suggesting that regardless of multiplicity the planets in these systems are drawn from the same underlying $\Q$ distribution. We find weak evidence ($p_{KS}=0.084, p_{AD}=0.045$) that 2-planet systems may be drawn from a different population than higher multiplicity systems. These marginal test statistics would be expected if some, but not all of the 2-planet systems belong to the same underlying physical population as the high multiplicity systems, but with some planets undetected due to observational biases. However, because some of our other complexity measures are undefined for systems with fewer than 3 planets, we do not investigate this point further.

\subsection{Monotonicity, $\mathcal{M}$}

Several prior studies have found that exoplanets are preferentially arranged with larger planets exterior to smaller planets \citep{Ciardi2013, Millholland2017, Weiss2018, Kipping2018}, although there has recently been some controversy surrounding this claim \review{\citep{Zhu2019, Murchikova}}.

In order to capture the degree to which planets are ordered by mass, we define the \textbf{monotonicity} as

\begin{equation}
    \mathcal{M} \equiv \rho_S \Q^{1/N}
\end{equation}

where $\rho_S$ is the Spearman rank-order coefficient calculated using planet masses as the input variables. The rank-order takes a value $\rho_S = 1$ for perfectly positive monotonic systems, $\rho_S = -1$ for perfectly negative monotonic systems, and $\rho_S=0$ for systems with no evidence of monotonic behavior. We include the multiplicative factor $\Q^{1/N}$ because while rank-order captures the degree to which a sequence is monotonic, it provides no information regarding the magnitude of any such monotonic trend. For example, the sequences [1, 3, 2] and [1, 100, 10] both have the same $\rho_S$ despite their obvious differences in scale. Multiplying by $\Q$ downweights $\mathcal{M}$ towards zero for any systems in which the planets are close to evenly sized and thus for which any evidence of monotonicity is more likely to be due to statistical noise. The factor $1/N$ scales this $\Q$ factor so that a high multiplicity system will have its evidence of monotonicity preserved even if there is little variation in planet masses. Using $\rho_S$ directly would also be inappropriate because it can only take a few discrete values for small $N$ and thus is not on its own flexible enough to describe the apparent size-ordering of planets. Note that $\mathcal{M}$ is normalized to the range (-1,1) and has the same qualitative interpretation as $\rho_S$.

\review{\citet{Kipping2018} performed an in-depth study of planet size ordering and argued that based on the solar system, consideration of size ordering should not be restricted to a single monotonic trend. Rather, because the Solar system planets (mostly) increase in size up to Jupiter and then decrease in size thereafter, a two-part trend best captures the high degree of apparent size ordering in the Solar system, whereas a one-part trend would indicate no size ordering. However, because the sample of planets considered in our present work is dominated by 3- and 4-planet systems, we restrict ourselves to a single component treatment of monotonicity. In the future, should sufficient very high multiplicity ($N \geq 5$) systems be discovered, the two component treatment advocated for by \citet{Kipping2018} should then be revisited.}

\begin{figure}
    \centering
    \includegraphics[width=0.45\textwidth]{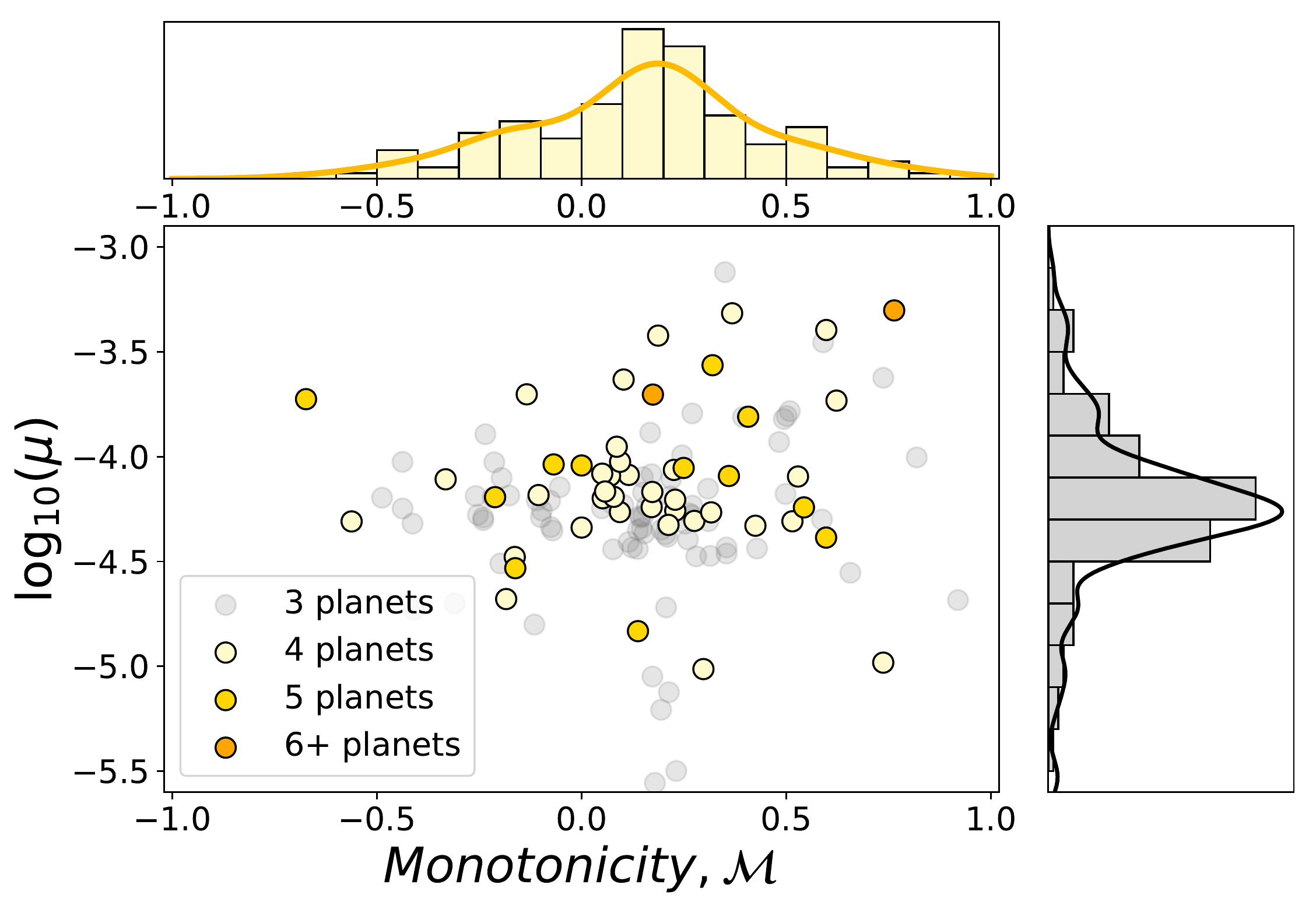}
    \caption{Distribution of monotonicity, $\mathcal{M}$, vs. dynamical mass $\mu$. Each 1-dimensional histogram is overplotted with a Gaussian KDE, the bandwidth for which was selected using Silverman's rule. The distribution of $\mathcal{M}$ is peaked at small positive values, inditing that there may be preference for planets to be arrange with larger bodies exterior to smaller bodies. At present, it is unclear whether this trend in monotonicity is physical in nature of the result of observational biases.}
    \label{fig:monotnoicity}
\end{figure}

We find that 72\% of high multiplicity ($N\geq3$) systems have $\mathcal{M}>0$, strengthening the finding of \citet{Weiss2018} that 65\% of planet pairs are ordered with the larger planet exterior to the smaller planet. In addition, while the smallest observed negative monotonicity value is $\mathcal{M}=-0.67$, there are five systems which have at least equally strong positive monotonicity $\mathcal{M} > +0.67$. Similarly, although there are only two systems with $\mathcal{M}<-0.5$, there are sixteen systems with $\mathcal{M}>+0.5$. In section \ref{sec:Synthetic} we investigate whether this apparent size ordering is physical in nature or the result of observational biases. The distribution of $\mathcal{M}$ is shown in Figure \ref{fig:monotnoicity}.

\subsection{Characteristic spacing, $\mathcal{S}$}

To describe the orbital spacing of a system, we define  the \textbf{characteristic spacing}, $\mathcal{S}$ as the average separation between planets in units of mutual Hill radii

\begin{equation}
    \mathcal{S} \equiv mean(\Delta_H)
\end{equation}

where the scaled separation, $\Delta_H$, and mutual Hill radius, $r_H$, for each adjacent planet pair are calculated from

\begin{equation}
    \Delta_H = (a'-a)/r_H
\end{equation}

\begin{equation}
    r_H = \left( \frac{m' + m}{3M_{\star}} \right)^{1/3} \left( \frac{a' + a}{2} \right)
\end{equation}

and $a$ is the semimajor axis of each planet \citep{Chambers1996}. Primed variables are outer planets in an adjacent pair and unprimed variables are inner planets.

\begin{figure}
    \centering
    \includegraphics[width=0.45\textwidth]{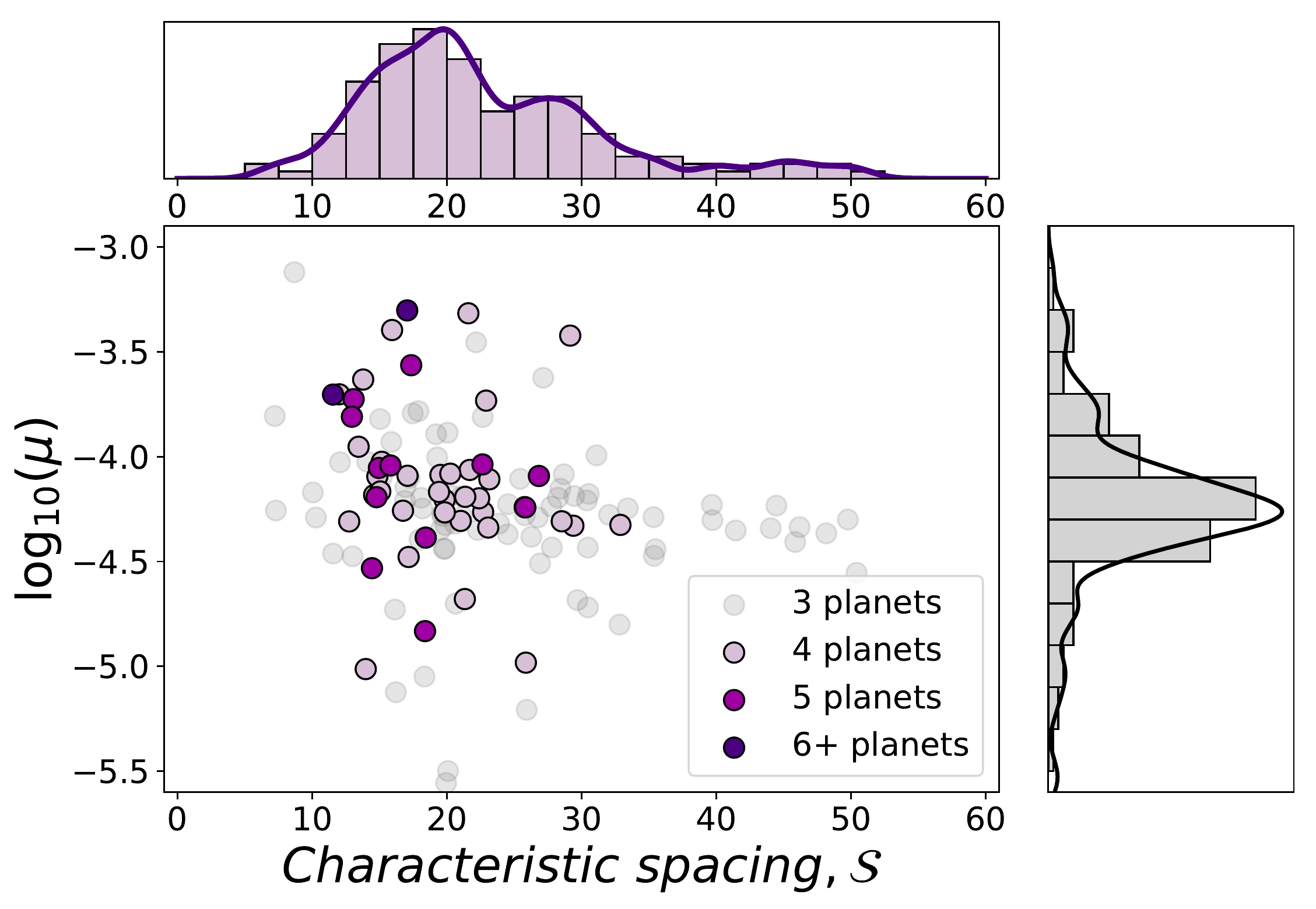}
    \caption{Distribution of characteristic spacing, $\mathcal{S}$, vs. dynamical mass $\mu$. Each 1-dimensional histogram for is overplotted with a Gaussian KDE. For $\mu$ the bandwidth was selected using Silverman's rule, whereas for $\mathcal{S}$ the bandwidth was selected using the improved Sheather-Jones algorithm \citep{SheatherJones1991, Botev2010}, which does not assume unimodality. The distribution of $\mathcal{S}$ is peaked at $\sim20$ mutual Hill radii, in agreement with \citet{Weiss2018}. We also find a tentative secondary peak near $S\approx30$. This ``echo'' peak would be expected if there is a large population of intrinsic 4-planet systems in which one of the intermediate planets has not yet been detected.}
    \label{fig:spacing}
\end{figure}

Figure \ref{fig:spacing} shows the behavior of $\mathcal{S}$ vs $\log\mu$. We reproduce previous findings \citep{Lissauer2011, Fang2012, PuWu2015, Dawson2016, Weiss2018} that planets are separated by \sim20 mutual Hill radii. We also identify a tentative secondary peak at $\mathcal{S}\approx28$. This ``echo peak'' would be expected near $\mathcal{S}\approx30$ if there exists a significant population of evenly-spaced 3-planet systems which are intrinsically 4-planet systems in which one of the two intermediate planets has not been detected. We discuss this hypothesized subpopulation further in our discussion of gap complexity and in Section \ref{sec:Trends} below. This result is in contrast to a previous result from \citet{Steffen2013} who found that based on period ratios, 2-planet systems and 4-planet systems are drawn from distinct populations. However, \citet{Steffen2013} considers a different population of systems (4 vs 2 planets) than we do here (4+ vs 3 planets). Based on K-S and A-D tests on $\mathcal{S}$, we do find that 2 planet systems are more widely spaced on average than 3 or 4+ planet systems, and so any tension between our results and those of \citet{Steffen2013} may turn out to be easily resolved.

\subsection{Gap complexity, $\C$}

We now define a measure which we term the \textbf{gap complexity}, following Equation \ref{C} and restated here as

\begin{equation}\label{C*}
    \C = -K \left(\sum_{i=1}^n p_i^*\log p_i^* \right)
    \cdot \left( \sum_{i=1}^n (p_i^* - \frac{1}{n})^2 \right)
\end{equation}

\begin{equation}\label{eq:p*}
    p_i^* \equiv \frac{\log(P'/P)}{\log(P_{max}/P_{min})}
\end{equation}

where $K$ is a normalization constant, $n=N-1$ is the number of gaps between planets and $p_i^*$ are pseudo-probabilities computed from the planets' orbital periods using Equation \ref{eq:p*}. Here, $P'$ is the orbital period of the outer planet in an adjacent pair, $P$ is that of the the inner planet, $P_{max}$ is the maximum period of any planet in the system, and $P_{min}$ is the minimum period. In this way $p_i^*$ are automatically normalized so that $\sum p_i^* = 1$.

Because orbital periods are not so easily converted to probabilities as masses are, gap complexity should be interpreted as a purely descriptive term and not as a traditional physical quantity. We do note, however, that orbital periods could in principle be transformed into specific orbital energies, which could then be re-expresed as occupancy state probabilities. Nevertheless, these extra steps would not necessarily result in a more comprehensible outcome, so in order to keep things simple and maintain focus on the ``big-picture'' of our methodology, we opt to work directly with orbital periods here.

The normalization constant, $K$, is chosen so that $\C$ is always in the range (0,1). A system with planets evenly spaced in log-period will have $\C=0$, while $\C \rightarrow 1$ when the maximum $p_i^* \approx 2/3$ \citep{Anteneodo1996}; the exact value depends on the number of planets in the system. $K$ can equivalently be expressed as $1/\C_{max}$, where $\C_{max}$ is the maximum possible complexity for a given $n$. \citet{Anteneodo1996} derive equations to numerically determine $\C_{max}$, and we present values for $n \leq 9$, corresponding to multiplicities for all known planetary systems, in Table \ref{tab:Cmax}. Alternatively, $\C_{max}$ can be approximated from the relation

\begin{equation}\label{eq:Cmax}
    \C_{max} \approx 0.262\ln(0.766n)
\end{equation}

\begin{figure}
    \centering
    \includegraphics[width=0.45\textwidth]{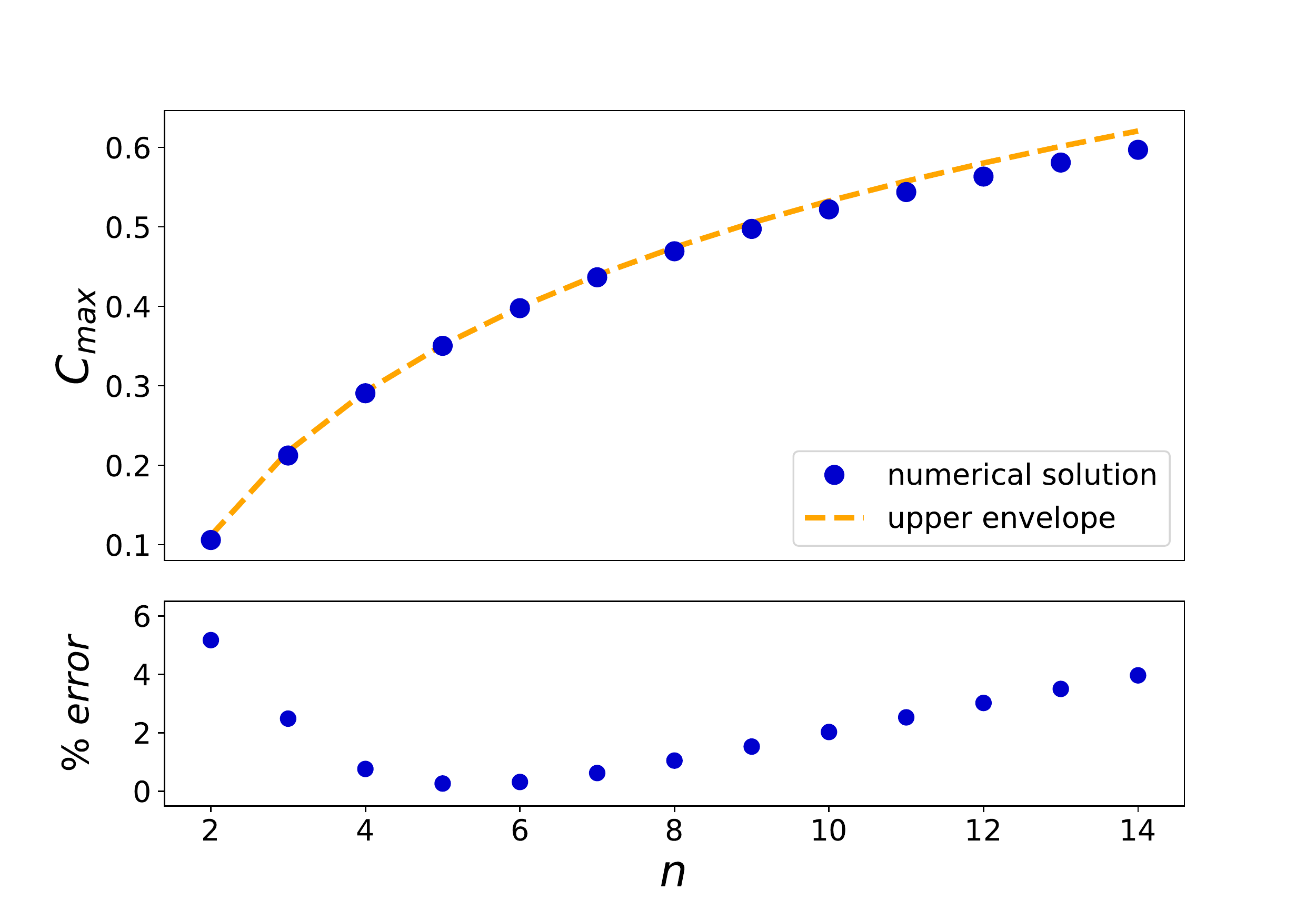}
    \caption{Numerically computed values of $\C_{max}$ for $n\leq14$, where $n = N - 1$ is the number of gaps between planets. The numerical values are well approximated by an upper envelope $\C_{max} \approx 0.262\ln(0.766n)$, which is accurate to within $6\%$ for the values of $n$ shown. Using an upper envelope ensures that when this approximation is used $\C_{max}$ will remain in the range (0,1).}
    \label{fig:Cmax}
\end{figure}

The above relation is determined by fitting a power law to the numerically determined maximum complexity, $\C_{max}$ vs $n$ for $n\leq9$. The power law fit is then shifted to establish an upper envelope so for $\C_{max}$ so that after normalization $\C$ remains in the range (0,1). We stress that Equation \ref{eq:Cmax} is provided merely for convenience and is an empirical quantity. In practice, using equation \ref{eq:Cmax} amounts to less than 2.5\% error for systems with $4\leq N\leq10$ and less than 6.1\% error for all systems with $N \leq 15$ (Figure \ref{fig:Cmax}). A python script for calculating $\C_{max}$ by numerically solving the equations of \citet{Anteneodo1996} will be made available on github\footnote{\texttt{https://github.com/gjgilbert/maiasaurus}}.

\begin{table}[tb]
\centering
\caption{Numerically determined values of $\C_{max}$ vs $n$. Note that the values shown are for the number of gaps, $n$, rather than the number of planets, $N = n+1$, in a system}
\begin{tabular}{c|c}\label{tab:Cmax}
     $n$ & $\mathcal{C}_{max}$ \\
     \hline
     2 & 0.106  \\
     3 & 0.212  \\
     4 & 0.291  \\
     5 & 0.350  \\
     6 & 0.398  \\
     7 & 0.437  \\
     8 & 0.469  \\
     9 & 0.497
\end{tabular}
\end{table}

Having defined our gap complexity measure, one might reasonably ask why not simply use $D$ or $H$ directly to characterize the spacing between planets? $H$ can be easily ruled out because planets in most real systems are roughly evenly spaced, and so are near maximum entropy, whether using $p^*$ or even if using $P'/P$ directly. So, $H$ is of little practical use when comparing real system spacings. The choice of $\C$ over $D$ is admittedly something of a subjective choice. We opt for $\C$ for two main reasons. First, for the population of observed systems, there is little actual dynamical range in $D$. In other words, $D$ is relatively insensitive to the scale of variations in period spacings for real planetary systems. Second, we find that using $\C$ more closely matches our intuitive sense of which systems are complex and which are simple; the straightforward quadratic dispersion relation of $D$ does not adequately capture the variety of system architectures observed by \Kepler.

\begin{figure}
    \centering
    \includegraphics[width=0.45\textwidth]{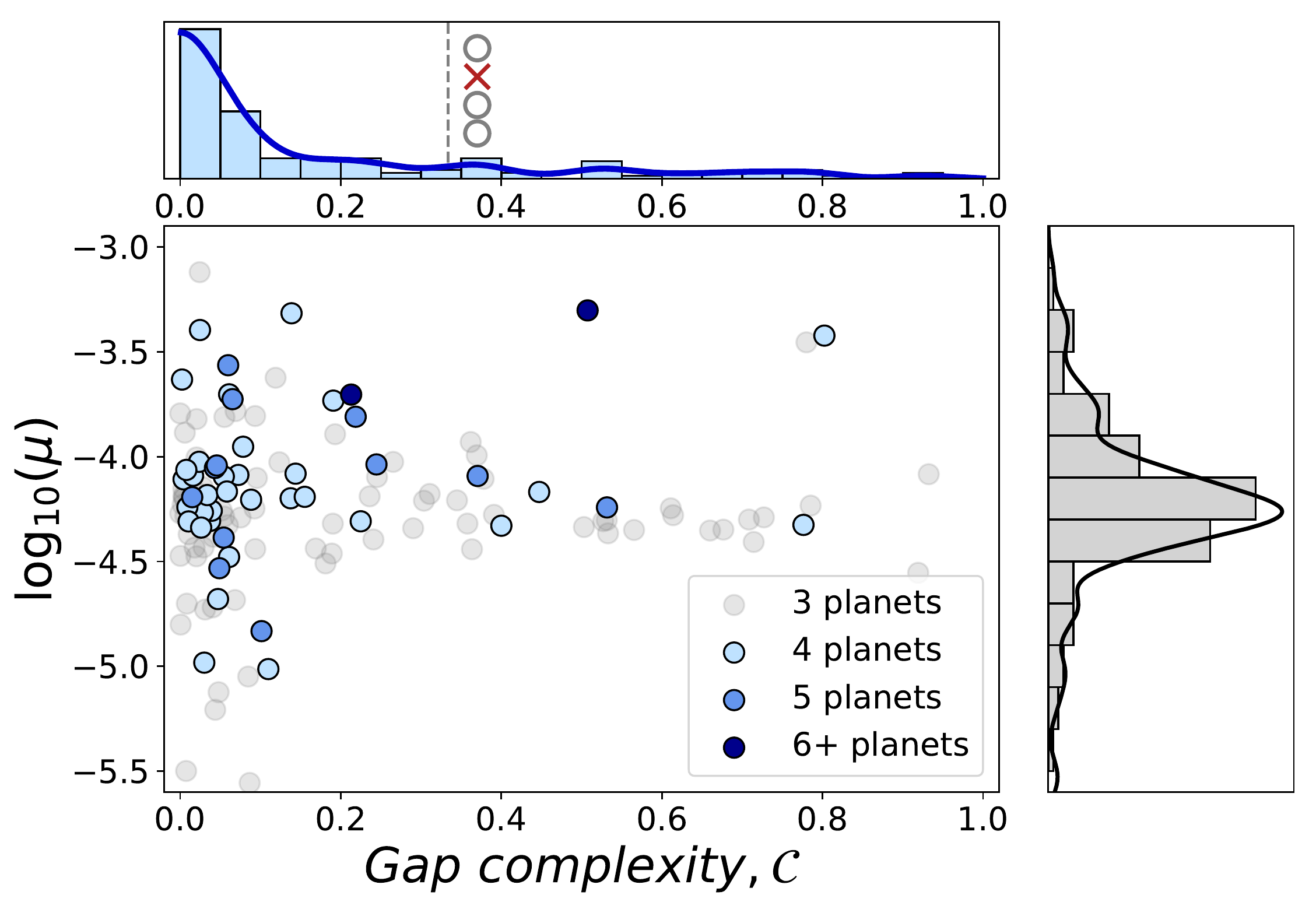}
    \caption{Distribution of gap complexity $\C$, vs. dynamical mass $\mu$. Each 1-dimensional histogram is overplotted with a Gaussian KDE, the bandwidth for which was selected using Silverman's rule. The majority of systems have $C \lesssim 0.2$, indicating that planets tend to be evenly spaced in log-period within a given system. There is a heavy tail extending out towards high values of $\C$. The majority of systems in this tail have dynamical masses consistent with the population mean, suggesting that this tail may be dominated by systems in which an intermediate planet has not yet been detected.}
    \label{fig:gapcomplexity}
\end{figure}

Like $\Q$, the distribution of $\C$ is peaked near zero, indicating that most systems are uniformly spaced, confirming the other primary result of \citet{Weiss2018}. In fact, when compared to population synthesis models, the system-level variable $\C$ indicates that planets within a system are even more regularly spaced than was previously inferred from pair statistics alone (see Section \ref{sec:Synthetic}). In addition to the peak at $\C=0$, we find a long, heavy tail containing some $\sim25\%$ of systems extending out to high gap complexities. One possible explanation for this high-complexity tail is that there exists a significant subpopulation of systems which host additional undetected planets at periods intermediate to the known planets. We explore this hypothesis in detail in Section \ref{sec:Trends}.

\subsection{Flatness, $f$}

To complete our classification scheme, we define a measure of \textbf{flatness}, $f$ that describes how close a system comes to that predicted for a completely ``cold'' architecture with circular and coplanar orbits. If such a system were completely edge-on to our line of sight, its transit durations would be perfectly predictable, equaling the orbital period times the ratio of the stellar diameter to the orbital circumference. If such a system were not edge-on, planets at larger semimajor axes would cut a smaller chord across the star and hence have a shorter transit duration. Converting semimajor axis to period using Newton's version of Kepler's third law, we have for the full (first to fourth contact) transit duration: 

\begin{equation}\label{eq:transit_duration}
    D = \Big{(} \frac{3P}{G \rho_\star \pi^2} \Big{)}^{1/3}  \bigg{(} (1+r)^2 - \Big{(}\frac{ G \rho_\star}{3 \pi} \Big{)}^{2/3}  P^{4/3} \cos^2 i \bigg{)}^{1/2}
\end{equation}
where $\rho_\star$ is the stellar mean density, $r=r_p/R_\star$, and $i$ is the inclination from the sky plane. We fit this function with weighted least-squares using only the free parameter $\cos i$.  The flatness measure $f$ is the sum of the squared residuals remaining after the fit, divided by the sum of the squared $D$ values before the fit.

We find that most systems are quite flat ($f$ near zero), in agreement with \citet{Fabrycky2014}. The definition of $f$ automatically normalizes to the range (0,1), but in practice we find very few systems with $f > 0.4$. Based on K-S and A-D tests, we find no evidence that 3-planet systems are drawn from from a different underlying flatness distribution than higher multiplicity ($N\geq4$) systems. This is in contrast to \citet{Zhu2018} who using additional input of TTV statistics found that higher multiplicity systems are intrinsically flatter. \review{We do find marginal evidence ($p_{KS}=0.043, p_{AD}=0.015$) that 2-planet systems are indeed flatter than 3-planet systems. Because the conclusions drawn by \citet{Zhu2018} are driven by 1- and 2-planet systems, perhaps any tension between our results and theirs will be easily resolved.}

\begin{figure}
    \centering
    \includegraphics[width=0.45\textwidth]{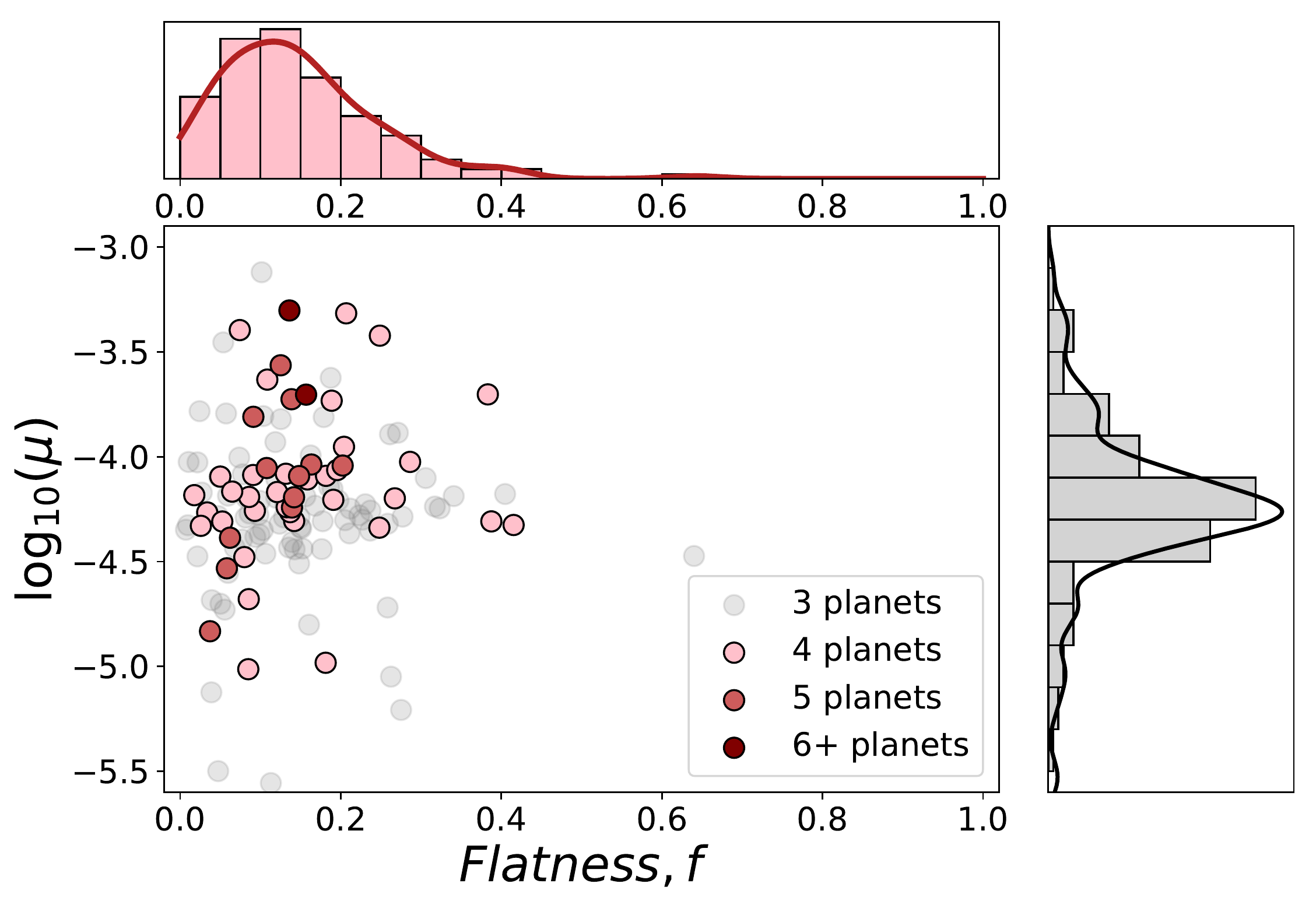}
    \caption{Distribution of flatness, $f$, vs. dynamical mass $\mu$. Each 1-dimensional histogram is overplotted with a Gaussian KDE, the bandwidth for which was selected using Silverman's rule. The majority of systems have small but non-zero flatness, and we find no obvious correlation between $f$ and $\mu$.}
    \label{fig:flatness}
\end{figure}

\section{Clustering and correlating}\label{sec:Clustering}

A primary benefit of our system-level approach is that we can now compare our measures against each other and search for correlations. Intuitively, one might expect that a system with equally-sized planets (low $\Q$) would also have planets which are equally-spaced (low $\C$) or tightly spaced (low $\mathcal{S}$), and a relatively coplanar geometry (low $f$). Interpreted through the lens of planet formation, it seems likely that a quiescent planet formation history that allows for equal mass partitioning might also allow for ordered period spacing and small mutual inclinations. Conversely, any chaotic stage during formation that disrupts orderly mass and energy partitioning would also be likely to excite inclinations.

In order to assess the strengths of relationships between our complexity measures, we employ the distance correlation metric \citep[dCor;][]{Szekely2007, Zucker2018}. Compared to the Pearson correlation coefficient, the distance correlation has the advantage of probing nonlinear relationships. Furthermore, dCor=$0$ only when two variables are independent. The value of dCor $\rightarrow 1$ for strongly correlated variables.

We find that the strongest relationships exist between $\Q$ and $\log\mu$ (dCor=$0.38$) and between $\C$ and $\mathcal{S}$ (dCor=$0.52$). These strong correlations are to be expected, as they represent the covariance between our two mass measures ($\Q, \mu$) and two spacing measures ($\C, \mathcal{S}$), respectively. We also find strong correlation (dCor=$0.48$) between $\Q$ and $\mathcal{M}$, which is expected from the definition of $\mathcal{M}$. The correlation between most other variable pairs is weak but nonzero ($0.1 <$ dCor $< 0.2$), except for between $\mathcal{S}$ and $f$ (dCor=$0.36$, $p<2\times10^{-4}$). These two variables are positively correlated, indicating that more tightly spaced systems are also flatter. \review{This finding bears out one of the main predictions of in situ planet formation models \citep{Dawson2016}.} The observed correlation could in principle be an imprint of the low eccentricities necessary for stability in tightly packed systems. However, because $f$ is more sensitive to inclinations than to eccentricities, we prefer the inclination interpretation. We do not find a commensurate correlation between $\C$ and $f$, but we do note that there is an observational bias toward low $f$. If a system were to have high mutual inclinations such that a planet is tilted off the limb of the star and not transit, our inferred value of $f$ would be lowered. In other words, there is a detection bias against highly inclined planets, which may produce a homogenizing effect on the observed flatness distribution. Hence, if the more unevenly spaced systems are indeed less flat, this effect would be difficult to observe directly, but it could be quantified with forward modeling.

Even though the two-variable correlations between most parameter pairs are weak, it may still be possible to uncover hidden structure by considering all variables simultaneously. To do so, we explore clustering in n-dimensional space by applying unsupervised clustering algorithms. The goal here is to determine whether compact \Kepler multis belong to one population or several, and, if more than one population exists, identify the key quantities which distinguish sub-populations of systems.

A similar endeavor was pursued by \citet{Alibert2019} who used t-distributed stochastic embedding \citep[T-SNE;][]{vanderMaaten2008} to automatically reduce the dimensionality of systems in ($R,P$) space. Our approach is complementary but has the advantage of searching for clusters in a more intuitive parameter space using quantities that are explicitly tied to each system's global architecture. The main difference between our approach and that of \citet{Alibert2019} is that T-SNE is a technique used for automatic dimensionality reduction of high-dimensional spaces, whereas our method is very distinctly hands-on. More specifically, \citet{Alibert2019} takes planetary radii and periods as input and then proceeds directly to a 2-dimensional index quantifying the similarity between exoplanetary systems. We have taken the intermediate step of first defining several quantities linked to the global architecture of the system. The choice of which approach to use may come down to a which is best for the objectives of a particular analysis. Because both T-SNE and unsupervised clustering require the definition of a distance metric, both techniques possess a similar subjective element. Ideally, both T-SNE and our cluster search should lead to the same conclusions regarding the architectures of multiplanet systems. We argue that using both methods provides a useful check.

We employ robust path-based spectral clustering \citep[R-PBSC;][]{Chang2007} in order to quantify the similarity between systems and search for possible subpopulations of systems. As a minimum criterion to sufficiently sample an n-dimensional space, at least $2^n$ samples are needed. With 129 systems, we can therefore consider up to seven dimensions. In practice, we use five: $\log_{10}\mu$, $\Q$, $\mathcal{S}$, $\C$, and $f$. We do not include $\mathcal{M}$ in our clustering searches because it is calculated explicitly from $\Q$ and is therefore correlated by definition. We do not include $N$ because the discrete nature of this variable leads clustering algorithms to simply group systems by multiplicity. 

Before proceeding, we apply multiplicative scaling factors to $\log_{10}\mu$ and $\mathcal{S}$ so that they are each approximately normalized to the range ($x, x+1$). This step weights each dimension equally in Euclidean distance space; $\Q$, $\C$, and $f$ are already normalized to this range by definition. We explore a range of normalization factors and find that our results remain qualitatively unchanged for $\log_{10}\mu \rightarrow \frac{1}{k_{\mu}}\log_{10}\mu$ if $1.5 < k_{\mu} < 4$ and for $\mathcal{S} \rightarrow \mathcal{S}/k_S$ if $30 < k_S < 80$. For the results presented here, we use $k_{\mu}=2$ and $k_S=40$.

The first step of standard spectral clustering is to construct a similarity matrix

\begin{equation}
    s_{ij}=
    \begin{cases}
			e^{\|x_i-x_j\|^2/2\sigma^2} & \text{$i \neq j$}\\
            0 & \text{$i = j$}
    \end{cases}
\end{equation}

which assigns a similarity value to each pair of points (i,j) based on their Euclidean distance in n-dimensional space. In context here, each point represents a single unique planetary system. The similarity matrix is mathematically equivalent to a graph representation in which each entry $s_{ij}$ is the edge strength between nodes. Path-based clustering modifies the similarity matrix to account for strong but indirect paths between nodes in the graph \citep{Chang2007}. A critical advantage of R-PBSC over standard spectral clustering is that is that the path-based approach is insensitive to choice of scaling parameter $\sigma$ which ordinarily can greatly impact the inferred clustering, yet is difficult to set in a self-consistent manner. The ``robust'' portion of R-PBSC refers to an additional weight $w_{ij}$ assigned to each pair of point in order to account for local variation of cluster sizes and densities. In particular, the data are weighted to prioritize paths that pass through dense regions but penalize paths which pass through sparse regions.

The common next step of all spectral clustering methods is to construct a graph Laplacian,

\begin{equation}
    L = s_{ij} - d_{ij}
\end{equation}

where

\begin{equation}
    d_{ij}=
    \begin{cases}
			\sum_j s_{ij} & \text{$i = j$}\\
            0 & \text{$i \neq j$}
    \end{cases}
\end{equation}

\begin{figure}
    \centering
    \includegraphics[width=0.45\textwidth]{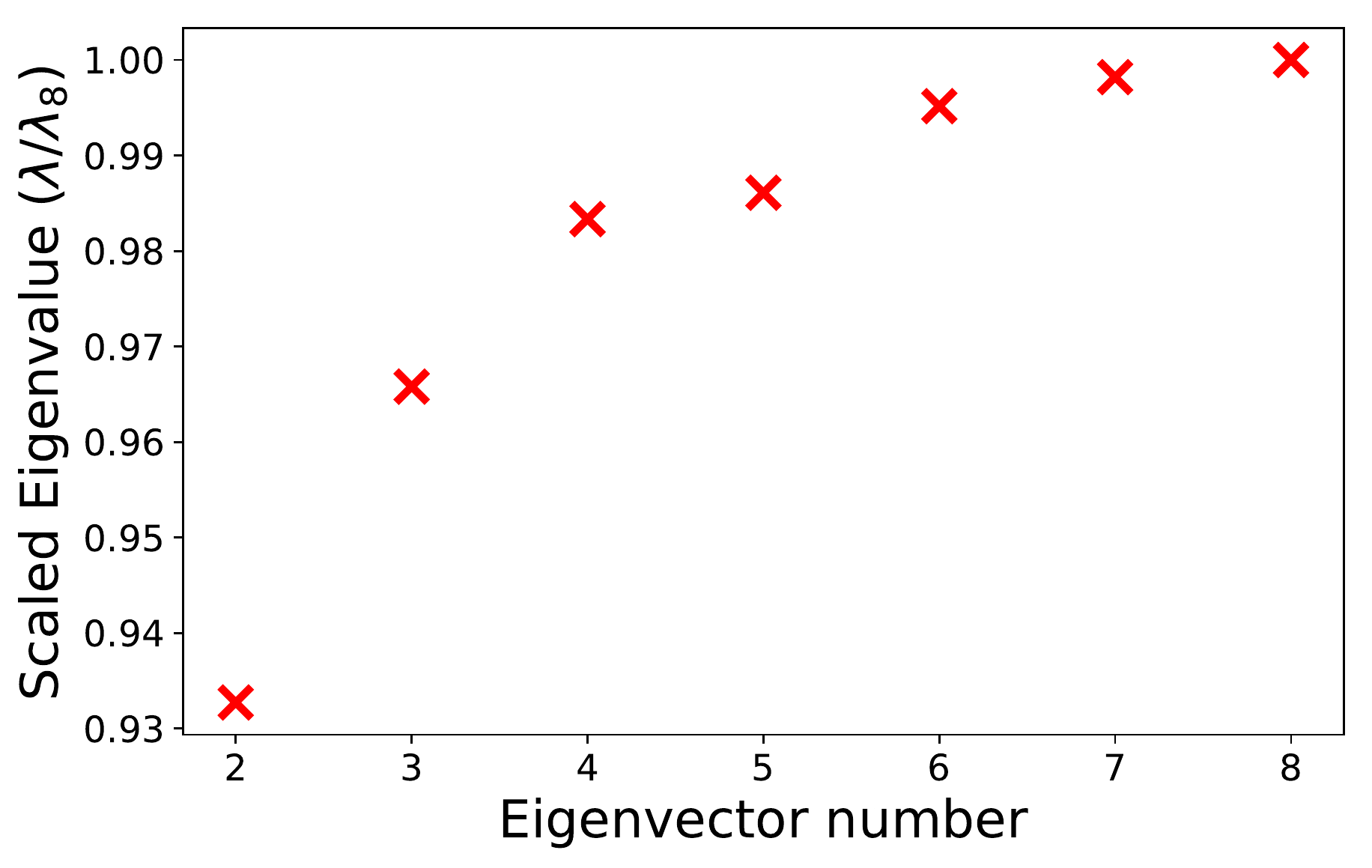}
    \caption{Eigenvalues for the 2nd-8th eigenvectors arising from our R-PBSC similarity matrix. The first eigenvalue is near zero ($\lambda_1 = 1.2\times10^{-16}$) and for clarity is not shown on the plot. The largest subsequent eigengap is between the 2nd and 3rd eigenvectors ($\lambda_2=0.938, \lambda_3=0.971$), indicating that the optimal number of clusters for the path-based solution is two \citep{Ng2001, vonLuxburg2007}. The eigengap $\lambda_3-\lambda_2$ is large compared to all other $\lambda_i-\lambda_{i-1}$, demonstrating that the clustering solution is stable.}
    \label{fig:eigengaps}
\end{figure}

Clustering is then performed on the first $k$ eigenvectors of $L$ using the basic k-means algorithm. In order to select the number of clusters, $k$, to consider, we employ the eigengap heuristic, which states that the clustering solution is most stable when $k$ is chosen such that the difference in eigenvalues is maximized \citep{Ng2001, vonLuxburg2007}. The \review{2nd-8th eigenvalues} for our clustering solution are shown in Figure \ref{fig:eigengaps}. In tests with synthetic data, we find that compared to standard spectral clustering, robust path-based clustering results in a clearer distinction between small eigengaps and large ones and therefore corresponds to more stable solutions.  Path-based clustering also more easily detects clusters with different numbers of members.

\begin{figure*}
    \centering
    \includegraphics[width=0.90\textwidth]{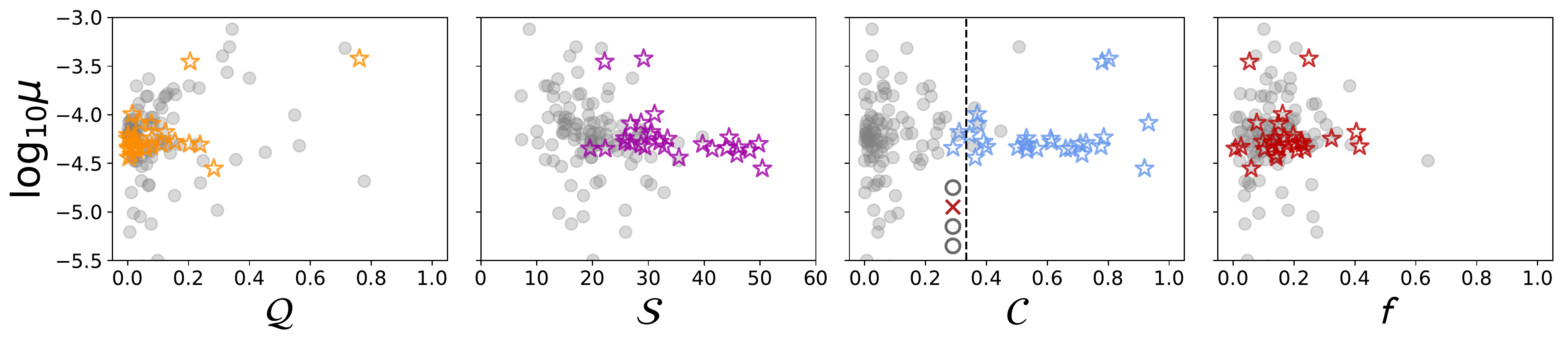}
    \caption{Results of robust path-based clustering assuming two clusters. The larger primary cluster (grey circles) centers around low $\Q$, $\C$, and $f$, with $\mathcal{S}\approx20$. The smaller secondary cluster (colorful stars) extends out to high values of $\C$ and $\mathcal{S}$. Of the 129 systems, 27 systems ($21\%$) fall in the smaller cluster. We interpret this secondary cluster as a population of systems in which an intermediate planet has not been detected. \review{The two outlier systems assigned to the secondary cluster are likely an artifact of the particular clustering method we employed which requires all data points to be assigned to one of the k=2 clusters. The astrophysical interpretation is that a small number of systems in our sample really are distinct from the main population in ways that cannot be attributed to observational biases.}}
    \label{fig:clusters}
\end{figure*}

We find that a 2-cluster solution is preferred (Figure \ref{fig:eigengaps}), with clusters that break fairly cleanly across the line $\C \approx 0.33$ (Figure \ref{fig:clusters}). Because this value of $\C$ corresponds to an architecture in which an evenly spaced ($\C=0$) 4-planet system has been reduced to a 3-planet system by removing one of the middle planets, an immediate explanation for this subcluster presents itself: these are systems which are missing an intermediate planet. Either the planet has not yet been detected or it does not exist. Because most of these high-$\C$ systems also have low $\Q$, and hence do not host a (known) giant planet that might have frustrated planetesimal growth, we deem the non-detection hypothesis more likely. We explore this hypothesis in detail in section \ref{sec:Trends} below.

The second largest eigengap occurs between the 3rd and 4th eigenvector, providing tentative support for a 3-cluster solution. We explore R-PBSC assuming 3 clusters and find that the results remain qualitatively unchanged except that the larger, low-$\C$ cluster is split into high-$\mu$ and low-$\mu$ partitions. The main clustering feature - a split at $\C \approx 0.33, \mathcal{S}\gtrsim30$ - remains, strengthening our belief that this high-complexity tail is a real feature of the population.
    
\review{To further validate our clustering stability, we perform two diagnostic tests. In the first test, we draw 64 samples without replacement from the population of systems and repeat our clustering analysis. In 18/20 iterations we find that a 2-cluster solution is indicated and we reproduce the dominant partitioning feature splitting the population of systems between low/high $\C$ and $\mathcal{S}$. We follow the same procedure drawing 81 and 112 samples and find good agreement with our initial results in 17/20 and 18/20 iterations respectively. In the few cases which fail to match our initial results, we find no clear distinction between large and small eigengaps, which we interpret as due to the fact that our clusters overlap and so any clustering results arise due to subtle differences in subpopulations which can become obscured if a few samples from the smaller secondary population are missing. In general, the stronger the indication towards a 2-cluster solution, the more cleanly the clusters separate along our initial partitioning. In the second test, we independently scramble each of the five dimensions ($\log_{10}\mu$, $\Q$, $\mathcal{S}$, $\C$, and $f$) and repeat our clustering analysis. Out of all 20 trials, none reproduced our original partitioning and in most cases there was no indication of any preferred eigengap at all. We interpret this test as an indication that clustering systems based on high/low $\C$ and $\mathcal{S}$ is a real feature and not an artifact of the data. For further background on clustering stability, we direct the interested reader to a review by \citet{vonLuxburg2010} and references therein.}

\review{One limitation of R-PBSC is that all data points must be assigned to one of the k=2 clusters without allowing for any outliers that do not belong to either of the groups. Indeed, there do appear to be a few points, mostly at large values of $\mu$ which fall at some distance from the main clump. The astrophysical interpretation is that a small number of systems in our sample really are distinct from the main population in ways that cannot be attributed to observational biases. In this particular case, a plausible explanation is that systems which manage to produce gas giants are atypical.}
    
\section{Comparison to synthetic catalogues}\label{sec:Synthetic}

There currently exist two state-of-the-art population synthesis models used for generating forward models of the \Kepler survey: the Exoplanet Population Observation Simulator \citep[EPOS;][]{Mulders2018,Mulders2019} and the Exoplanet System Simulator \citep[SysSim;][]{Hsu2018, Hsu2019, He2019}. Both models first generate a \textit{physical} population of stars and planets using realistic parameter distributions and then apply geometric and instrumental detection biases in order to simulate an \textit{observed} catalogue that would have been seen by \textit{Kepler}. In order to assess the physical population of systems underlying our observed trends, we compare our real CKS data to synthetic populations generated using each of these two simulators.

Because our complexity measures are almost purely descriptive, they should automatically match for real and synthetic data if the population synthesis models are accurate and complete. The most important physics underlying our scheme is the application of a mass-radius-period relation, which was applied uniformly to both real and synthetic systems. Where we have made other physical assumptions, notably for calculating mutual Hill radii or relating transit durations to inclinations - dependencies on planetary masses are weak and thus the particular choice of mass-radius-period relation is minimized.

In this section (\S\ref{sec:Synthetic}), we describe how EPOS and SysSim, as well as a directly bootstrapped catalogue, compare to the real data. We present an interpretation of significant trends in section \ref{sec:Trends}.

\subsection{EPOS}

We compare our real systems to a synthetic catalogue generated using EPOS. The catalogue was generated using multi-planet mode with all input parameters set to the optimized values found by \citet{Mulders2018}, except for the radius ratio distribution $r_p'/r_p$, which was modified to draw from a a log-normal distribution with dispersion 0.15 dex and mean of unity. All stellar hosts were identical to the Sun, i.e. $R_{\star} = R_{\odot}$, $M_{\star} = M_{\odot}$.

To ensure a valid domain for comparison, we restrict both the real catalogue and the synthetic catalogue to the radius and period limits used by \citet{Mulders2018} to optimize the EPOS fit parameters. We first eliminate any small planets ($r_p < 0.5 R_{\oplus}$) or long-period planets ($P > 400$ d), which is roughly analogous to setting more stringent detection thresholds. We then discard any systems which have been reduced to only one planet. Finally, we remove any systems which host at least one planet with $r_p > 6 R_{\oplus}$ or $P < 2$ d. The reason we eliminate entire systems rather than removing only the offending individual objects is because these large-radius or short-period planets are likely to be the most detectable objects and therefore correspond to artificial limitations of the simulator rather than natural observational bias.

We next recompute each of our complexity measures on the reduced collection of systems. Measures for the synthetic systems were computed following an identical procedure as for the real systems. In practice, this means that we we convert radii and periods to masses following \citet{Neil2019} and we calculate flatness from transit durations following Equation \ref{eq:transit_duration}.

\begin{figure*}
    \centering
    \includegraphics[width=0.90\textwidth]{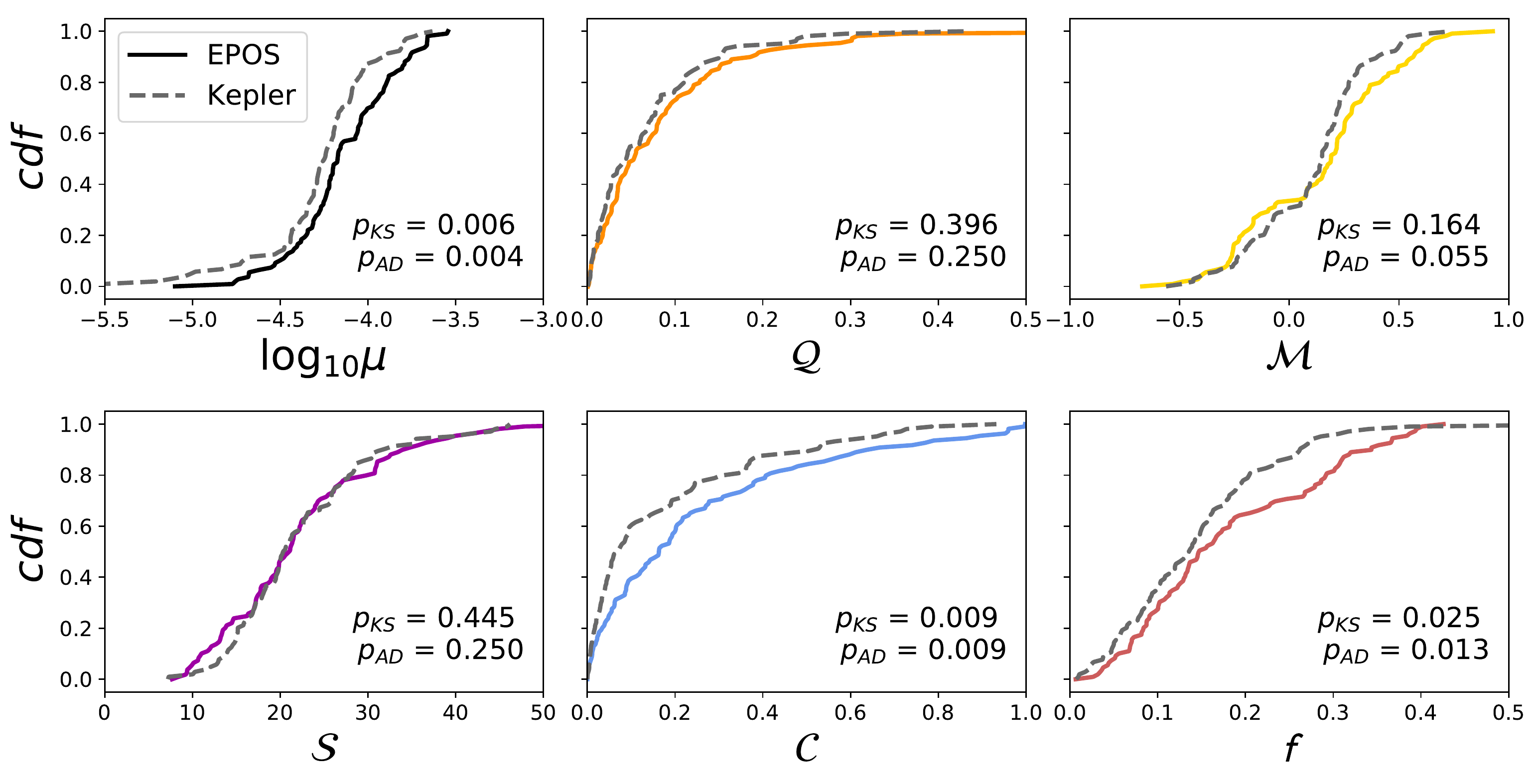}
    \caption{Cumulative density functions of system-level complexity measures for \Kepler compared to EPOS. Dashed grey lines give results for \Kepler and solid colored lines give results for EPOS. Resultant p-values of Anderson-Darling and Kolmogorov-Smirnov tests are shown on the plots. The data are well matched for $\Q$ and $\mathcal{S}$ but show statistically significant differences for $\log\mu$, $\C$, and $f$. It is unclear whether the underlying monotonicity distributions are different for the real vs. synthetic catalogues. Note that each panel has a different scale on the horizontal axis.}
    \label{fig:EPOS_cdf}
\end{figure*}

We find that the CKS and EPOS populations are well matched for 3 out of 7 of our measures (see Figure \ref{fig:EPOS_cdf}). Based on AD and KS tests, the (observed) real and synthetic populations are drawn from the same underlying distributions of mass partitioning $\Q$, characteristic spacing $\mathcal{S}$, and multiplicity $N$. However, the EPOS systems are slightly - but significantly - more massive than the CKS systems, as traced by dynamical mass $\log\mu$. In addition, the real systems are more evenly spaced than predicted by EPOS, as traced by gap complexity $\C$. The evidence as to whether the real and synthetic populations are drawn from different monotonicity distributions is unclear, although we do note that the \Kepler systems are slightly more concentrated towards $\mathcal{M}=0$ than the EPOS systems. We find strong statistical evidence that the two populations are drawn from distinct flatness distributions, but because our measure of $f$ depends on stellar properties, which were assumed to be uniform for EPOS, we hesitate to draw any definitive conclusions regarding the mutual inclinations of the systems.

\subsection{SysSim}

We compare our real catalogue to a synthetic catalogue generated using SysSim. The catalogue was generated with input parameters set to the optimized values found by \citet{He2019}. Our procedure for reconciling the limits of the simulated catalogue and the real CKS catalogue is identical to the procedure described for EPOS in the preceding section, although the bounds on period ($3 < P/$days $< 300$) and radius ($0.5 < r_p/R_{\oplus} < 10$) are slightly different. The procedures used to convert radii to masses and transit durations to flatness are likewise identical to the procedures described above.

\begin{figure*}
    \centering
    \includegraphics[width=0.90\textwidth]{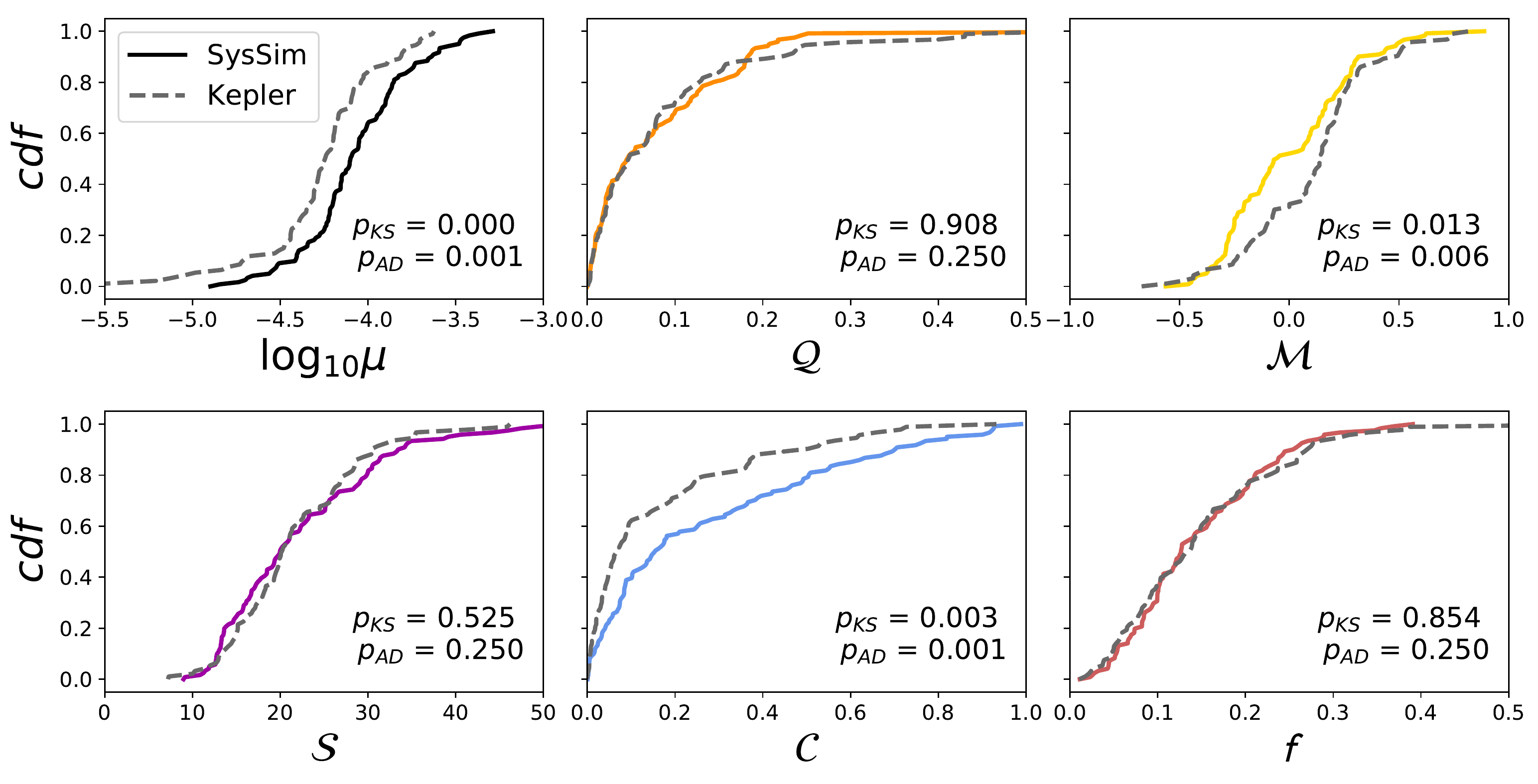}
    \caption{Cumulative density functions of system-level complexity measures for \Kepler compared to SysSim. Dashed grey lines give results for \Kepler and solid colored lines give results for SysSim. Resultant p-values of Anderson-Darling and Kolmogorov-Smirnov tests are shown on the plots. The data are well matched for $\Q$ and somewhat matched for $\mathcal{S}$ but show statistically significant differences for $\log\mu$, $\C$, $f$, and $\mathcal{M}$. Note that each panel has a different scale on the horizontal axis.}
    \label{fig:SysSim_cdf}
\end{figure*}

We find that the CKS and SysSim populations are well matched for $\Q$, $f$, $\mathcal{S}$, \review{and $N$} but show statistically significant differences for $\log\mu$, $\C$, and $\mathcal{M}$ (Figure \ref{fig:SysSim_cdf}). As with EPOS, SysSim over-predicts total system dynamical masses and under predicts the degree to which systems are clustered near very low gap complexity ($\C \rightarrow 0$).  SysSim, however, produces a population of systems which much more closely matches the flatness distribution of real systems. \review{The key difference between EPOS and SysSim is that while EPOS draws mutual inclinations from a single Rayleigh distribution, SysSim allows for two populations characterized by low and high mutual inclinations, each defined by a distinct Rayleigh distribution. So, perhaps it is unsurprising that the more flexible model achieves a better fit to the data.} Given that SysSim also allows for a range of stellar densities, a key parameter for computing flatness, we suggest that incorporating more rigorous constraints on stellar masses and radii into future population synthesis models may be a fruitful avenue for exploration.

The tension for monotonicity, $\mathcal{M}$, between \Kepler and SysSim is unexpected, and so we investigate it in greater detail. In particular, SysSim produces an overabundance of systems with low negative monotonicities. We recalculated $\mathcal{M}$ directly using planetary radii without converting to mass first, but this did not bring the model and data into agreement, indicating that the unexpected monotonicity relation is not an artifact of our mass-radius-period relation. Furthermore, the monotonicity distribution remained qualitatively unchanged for several generations of the SysSim catalogue, demonstrating that these results are robust and not a statistical fluke. The tension between the data and SysSim outputs may simply be due to the fact that we have employed SysSim only to provide a point estimate for comparison based on previously optimized inputs, whereas the motivation behind SysSim is to generate a distribution of models. Conditioning the outputs of SysSim on our newly defined measures and producing a suite of simulations may automatically resolve this tension. However, modification of forward models is beyond the scope of this paper, and so we leave such a project for future work.

\subsection{Direct downsampling}

In addition to comparing against forward models, we also compare the data against itself by directly downsampling the highest multiplicity systems ($N\geq4$) to generate a synthetic population of 3-planet systems. To produce this population, we randomly draw - with replacement - one of the 4+ planet systems from the CKS catalogue. We then generate a random number between 0 and 1 for each of the planets in the system, discarding any individual planets whose geometric transit probability ($P_{transit} \sim 1/a$) is less than the random number. If the downsampled system contains exactly 3 planets, we add it to our downsampled catalogue and repeat this procedure until the number of systems in the downsampled catalogue matches the number of 3-planet systems in our CKS catalogue. This procedure assumes moderate-to-high mutual inclinations between the planets and is admittedly less sophisticated than the routines run by EPOS and SysSim. Nevertheless, directly comparing the data against itself allows us to investigate whether the 3 planet and 4+ planet systems are drawn from the same intrinsic physical distribution.

\begin{figure*}
    \centering
    \includegraphics[width=0.90\textwidth]{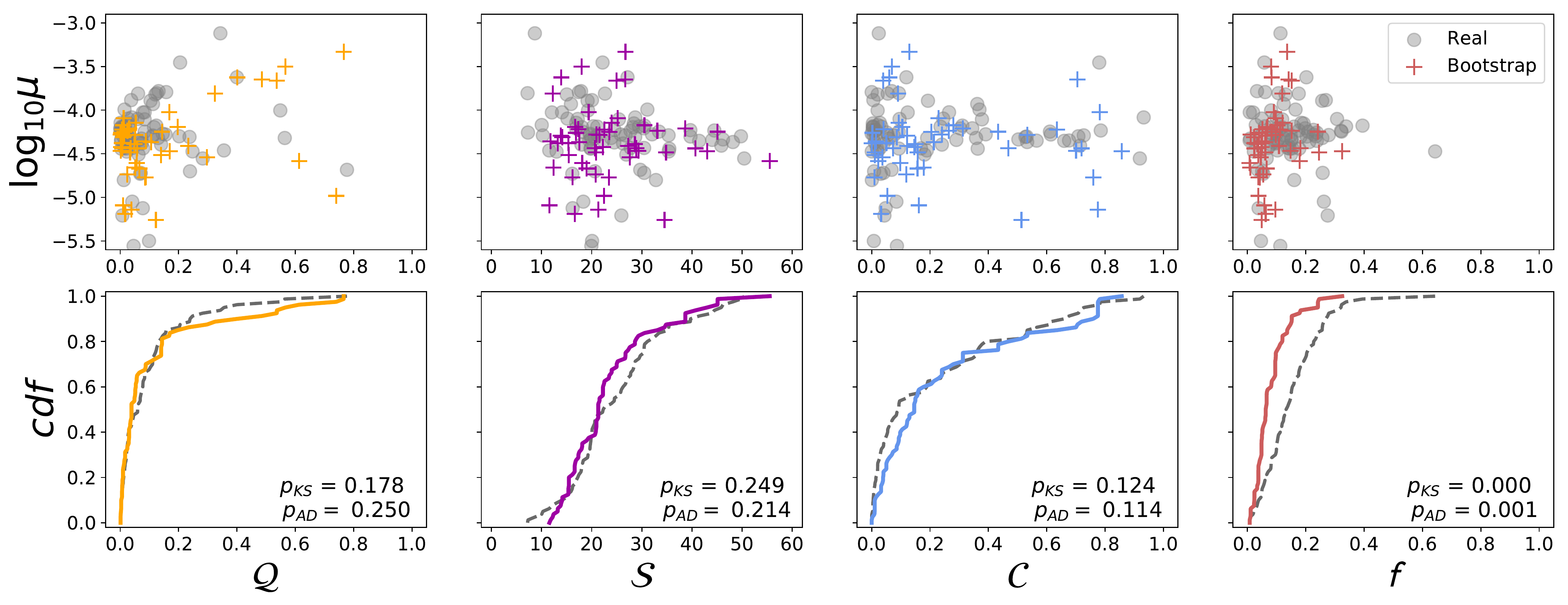}
    \caption{Downsampled bootstrap 3-planet systems compared to the real 3-planet systems from CKS. This is for a single iteration, but the results are representative of a typical outcome}
    \label{fig:Downsampled_3s}
\end{figure*}

The results of our downsample comparison are shown in Figure \ref{fig:Downsampled_3s}. We find that the downsampled 3 planet systems and real 3 planet systems are closely matched in $\Q$, $\mathcal{S}$, and $\C$ but that the downsampled systems are significantly flatter than the real systems. Notably, direct downsampling reproduces the distribution of $\C$ values much better than either of the population synthesis models. We interpret these results as evidence that all of the systems in our sample - regardless of multiplicity - are drawn from the same underlying physical distribution and that inter-system differences based on multiplicity can largely be explained by detection selection effects.

\section{Discussion}\label{sec:Trends}

Comparison of our system level measures between real CKS data and simulated catalogues from EPOS and SysSim provokes many questions about the architectures of exoplanetary systems. Below we discuss several such questions.

\subsection{How strong is the trend toward uniform intra-system sizing and spacing?}

\review{The question of whether planets in a system are ``peas in a pod'' has been a subject of intense debate in recent years, with several studies \citep{Millholland2017, Wang2017, Weiss2018} finding evidence for preferential arrangements of sizes and orbital periods, whereas others \citep{Zhu2019, Murchikova} have argued that these correlations are due to selection effects. There are three questions being debated here: are planets within a system preferentially (1) the same size, (2) evenly spaced, or (3) ordered with larger planets exterior to smaller planets. We can summarize our response to these three questions as (1) probably, (2) yes, and (3) unclear. The key advantage of our analysis over prior works is that we arrive at our conclusions by comparing \Kepler data directly to forward models (EPOS, SysSim) rather than employing bootstrap tests. This approach is in fact advocated for by \citet{Zhu2019} which identifies forward modeling as the best means to resolve questions concerning planetary system architectures.}

\review{Both EPOS and SysSim match the mass partitioning ($\Q$) distribution from CKS to a high degree of fidelity confirming the primary conclusions of \citet{Wang2017}, \citet{Millholland2017}, and \cite{Weiss2018} that planets within a system tend to be the same size. Our results contradict the hypothesis of \citet{Zhu2019} and \citet{Murchikova} that intra-system correlations are due to selection effects, although we do agree with their critique that the Pearson correlation coefficient is insufficient to capture the observed trends. The trend towards uniformity had previously been supported by \citet{He2019} and \citet{Sandford2019}. By employing two sophisticated forward models which explicitly assume correlations in planet size and combining these models with a system-level analysis, we find strong evidence that the majority of \Kepler high-multiplicity systems host planets which are remarkably similar in size ($\Q\rightarrow0$).}

Our analysis of system spacing strongly supports the conclusion that planets within a system tend to be evenly spaced ($\C \rightarrow 0$) and separated by approximately 20 mutual Hill radii ($\mathcal{S} \sim 20$). In fact, based on our system-level analysis of gap complexity, we find that planets within a system are even more uniformly spaced (in log-period) than predicted by either forward model. That is, the distribution of $\C$ is more strongly peaked towards zero for real systems than for synthetic systems. This is not to imply that the models are insufficient to describe real planetary systems, but rather employing settings that had been optimized based on pair statistics (i.e. period ratios) does not fully capture the higher-order patterns present in real systems. We see this as a major success of our method and a clear illustration that a system-level analysis is both warranted and necessary to understand the architectures of multiplanet systems.

The evidence for preferential size-ordering, however, is mixed. Although the distribution of $\mathcal{M}$ matches reasonably well between data and EPOS, SysSim produces too many stems with low or negative monotonicity. Neither forward model explicitly assumes any size ordering, so it is difficult to say what the correct interpretation is. Given that so many systems host planets which are all essentially the same size ($\Q\approx0$), the simplest explanation is that for many system any evidence of monotonic ordering may largely be due to random noise. Removing the lowest $\Q$ systems may reveal hidden trends in monotonicity. However, given the low number of high-$\Q$ systems in our sample, there are not enough systems to explore this idea here. Applying this method to the full \Kepler sample may give us enough data. This is an advantage (and a possible test) of our method - whereas it would be nearly impossible to say which individual planets to remove, it is now possible to identify systems to remove in order to seek out more subtle trends. \review{It is also possible that size ordering of planets is a real effect only for planet pairs - for example, pairs straddling the photoevaporation valley - but not on a whole system level. There may be several astrophysical processes in action which when overlaid obscure one another. Further work is needed to resolve these questions.}

\subsection{Why do forward models over-predict inferred system dynamical masses?}

Both forward models produce a population of planetary systems with dynamical masses, $\mu$, slightly but significantly higher than those of the real CKS systems. We offer several explanations to explain this tension.

First, we did not refit the forward models to match the data, but rather used previously optimized inputs based on \citet{Mulders2018} for EPOS and \citet{He2019} for SysSim. These prior results considered many 1- and 2-planet systems which were not studied here, and so the discrepancy could naturally arise if the planets in the high-multiplicity and low-multiplicity systems are drawn from distinct mass distributions, or if the planets are drawn from different regions of the same underlying distribution. In some sense we have not used the forward models exactly as intended (i.e. producing a suite of models to interactively fit to the data), and so some tension is to be expected. Such a project is beyond the scope of this work, but we emphasize that even ``out-of-the-box'' both forward models perform remarkably well along most system-level dimensions.

Second, we have employed a new joint mass-period-radius relation \citep{Neil2019}. Because the forward models were developed to match planetary radii and not planetary mass, some discrepancy when working with mass is to be expected. \review{SysSim estimates masses using a different mass-radius relation than we do \citep{Ning2018}, but because any planetary mass dependence in SysSim comes into play only in the context of dynamical stability tests, we expect any differences which arise due to choice of mass-radius relation to be slight.} However, because (dynamical) mass is more fundamentally tied to planet formation, we stand by our decision to work in mass space rather than in radius space. \review{With a number of mass-radius relations currently in use \citep{Weiss2014, Wolfgang2016, Chen2017, Ning2018, Neil2019}, we suggest that an in-depth direct comparison of these various mass-radius relationships might be performed in order to allow greater confidence and precision when interpreting results produced using different relations.}  We do not see the tensions we have identified as a problem for either the mass-radius relations or for the forward models, but rather an indication that more work is needed to combine these features into a uniform analysis.

Third, neither forward model yet incorporates \review{a fully self-consistent model of stellar properties}. In fact, EPOS assumes that all stars are stellar twins. \review{Although SysSim draws stellar radii and masses from high precision, reliable catalogues provided by Gaia DR2, at the time of our analysis, the software did not yet impose a joint constraint on stellar density or explicit correlations between stellar type and planet properties, leading to a few spurious outliers. Investigation of planet-star correlations is an active field of study, and between the time of our analysis and initial submission of this paper, the SysSim group has already published an update which treats stellar properties in a more sophisticated manner \citep{He2020} which addresses many of our concerns.} Dynamical mass is explicitly linked to stellar mass, and so it would be unreasonable to expect a perfect match between models and data \review{before the models had implemented a detailed treatment of stellar host properties. We look forward to ongoing collaboration with both the EPOS and SysSim teams in the future.}

\subsection{What is the true distribution of mutual inclinations?}

All three of our synthetic catalogues lead to qualitatively difference conclusions regarding the flatness distributions of the real \Kepler systems. EPOS underpredicts the observed flatness, direct bootstrapping overpredicts the observed flatness, and SysSim gets the flatness distribution just about right. One interpretation is that SysSim has accurately captured the stellar, planet, and instrument properties that lead to the observed system architectures. Another interpretation is that although the central trend in mutual inclinations - the nearness of exoplanetary systems to coplanarity - remains undisputed, we do not yet understand the details of the distribution. Until the differences between methods can be brought into agreement, it will be difficult to say which interpretation is correct. Because inclination in the property most closely tied to geometric detection biases, understanding the details of system coplanarity is of the utmost importance for determining the true architectures, and thus formation histories, of exoplanetary systems.

\subsection{Can missing planets explain the observed system-level trends?}

\begin{figure}
    \centering
    \includegraphics[width=0.45\textwidth]{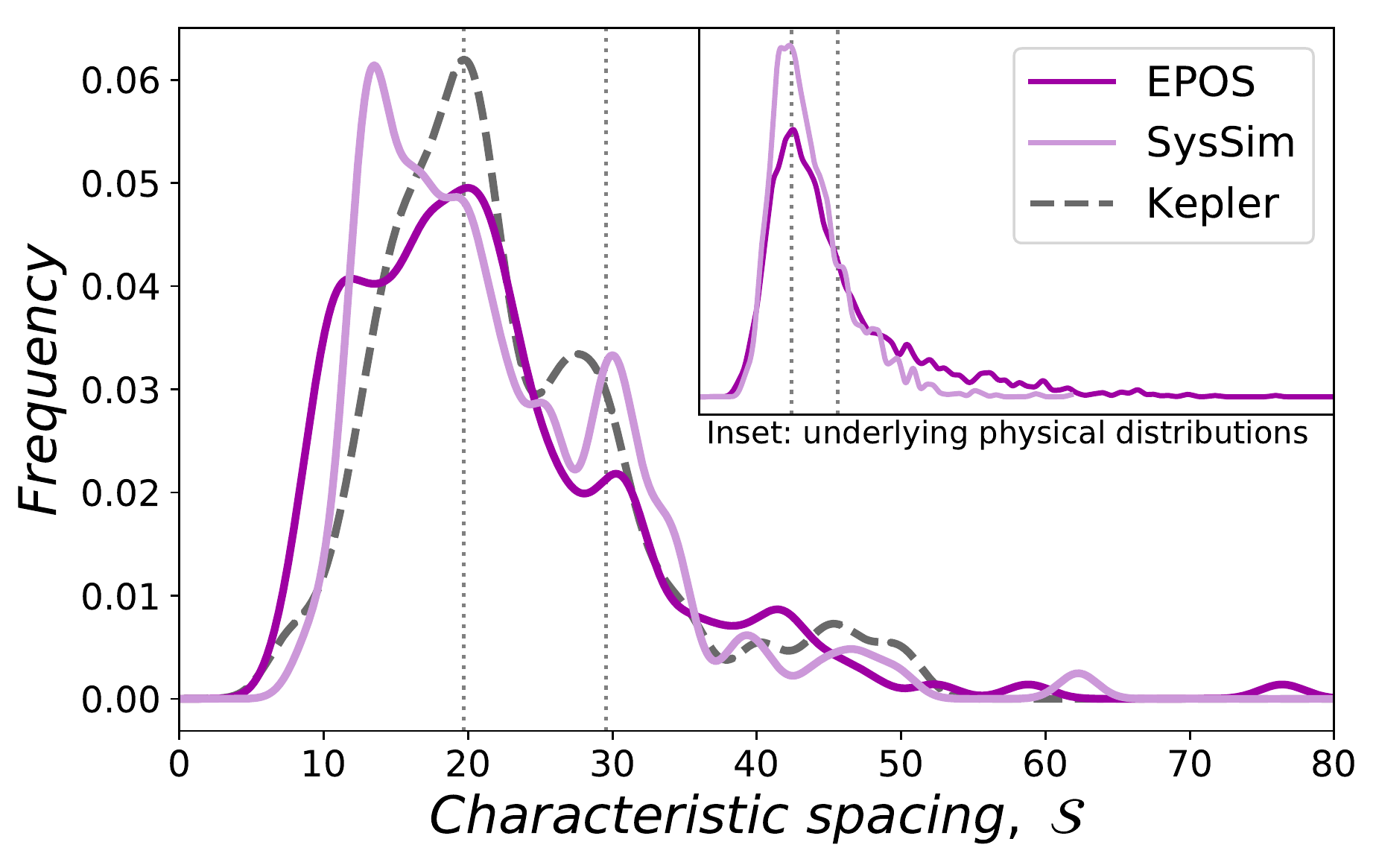}
    \caption{Gaussian kernel density estimate for the occurrence frequency of characteristic spacing, $\mathcal{S}$, scaled to the peak of the distribution. The distributions for \Kepler (grey dashed line), EPOS (dark purple) and SysSim (light purple) are all strongly peaked near $\mathcal{S}\approx20$, with a more subtle secondary peak at $\sim1.5\times$ this value. This ``echo peak'' would be expected if there exists a significant population of intrinsic 4-planet systems in which one of the two intermediate planets has not yet been detected. Indeed, the presence of such a peak at this location in all three datasets is surprising unless there are missing planets. The fluctuations at low $\mathcal{S}$ near the peaks of the distributions from EPOS and SysSim are not statistially significant (see Figures \ref{fig:EPOS_cdf} \& \ref{fig:SysSim_cdf}).}
    \label{fig:echo_peaks}
\end{figure}

The observed system-level trends can be most easily understood if all (or at least most) \Kepler high-multiplicity systems are drawn from the same intrinsic distribution with a sub-population of systems ($\sim20\%$) hosting additional undetected planets intermediate to the known planets. The evidence for this scenario is as follows. First, unsupervised clustering finds a sub-population of systems at high C and S (Figure \ref{fig:clusters}). Of the 129 systems, 27 systems (21\%) fall in the smaller cluster. Second, we observe an ``echo peak'' in the distribution of $\mathcal{S}$ at $\sim30$, or 1.5 times the primary peak location. This ``echo peak'' is seen in both the real data and in population synthesis models, but critically is not present in the underlying physical distributions of the models (Figure \ref{fig:echo_peaks}). Such an ``echo peak'' is to be expected if there exists a large population of intrinsic 4-planet systems in which one of the two intermediate planets has not yet been detected. Third, the distribution of $\mu$ is matched after normalizing by multiplicity (Figure \ref{fig:mu_cdf}), indicating the typical planet mass is the same regardless of multiplicity. Although these trends could conceivably be explained but multiple physical populations of systems, because the trends can be readily explained by a single physical population convolved with known observational biases, we deem the single population explanation most likely.

In addition, the distribution of $f$, while not direct evidence of missing planets, is also consistent with this hypothesis. Naively, one might expect that lower multiplicity systems might have greater scatter and thus larger $f$ than higher multiplicity systems. However, comparison of 3 and 4+ planet systems via KS and AD tests on $f$ reveal no distinction between these two populations. Because it is the most highly inclined planets that will be missed, there is thus a homogenizing effect on $f$, and therefore a lack of distinction in $f$ between observed populations is not necessarily an indication of a lack of distinction between underlying physical populations. An alternative way forward is to incorporate the hypothesis of \citet{Zhu2018} that higher intrinsic multiplicities are flatter into a forward-modelling approach that compares to our measures rather than only 2-planet measures. More work is needed to uncover the true distribution of mutual inclinations. At present, we can say that the observed distribution of $f$ is consistent with our hypothesis of missing planets.

Our explanation for the high-$\C$ subcluster makes the straightforward, testable prediction that these systems should host additional planets in their gaps. The evidence for such planets may already be present in the data, either as low signal-to-noise grazing transits or as dynamical signatures on the known planets that may be resolved via transit timing variations. In either case, because we now know where to look, it may be possible to identify marginal signals that would not pass muster in a blind search, but which - when placed in the context of their host systems - rise to the level of statistical significance. A search for these missing planets is beyond the scope of this paper, and we leave this search for future work. A similar project has previously been suggested based on a generalized Titius-Bode law \citep{Bovaird2013, Huang2014, Bovaird2015}, although the generality of Titius-Bode and its utility for planet searches remains questionable.

\section{Summary and Outlook}\label{sec:Summary}
We have proposed seven dimensions along which to characterize exoplanetary systems: multiplicity, $N$; dynamical mass, $\mu$; mass partitioning, $\Q$; monotonicity, $\mathcal{M}$; characteristic spacing, $\mathcal{S}$; gap complexity, $\C$; and flatness, $f$. We analyze these quantities in concert in order to search for system-level trends and identify sub-populations of exoplanetary systems. Our key findings are as follows.

\begin{itemize}
    \item Within a given system, planets tend to be the same size (low $\Q$), in agreement with prior studies.
    \item Within a given system, planets tend to be uniformly spaced in log-period (low $\C$). In fact, the degree of intra-system uniformity in spacing is greater than suggested by an uncorrelated set of planet pairs.
    \item Our method did not detect whether are preferentially ordered by size.
    \item Comparisons between forward models (EPOS and SysSim) and real data (CKS) suggest that current models have successfully captured the mass partitioning ($\Q$) but not the mass scale ($\mu$) of high-multiplicity systems; conversely, these models have successfully captured the period spacing scale ($\mathcal{S}$) but not the low spacing complexity ($\C$) of high-multiplicity systems.
    \item The observed trends in $\mu$, $\C$, and $\mathcal{S}$ can most readily be explained if all (or at least most) \Kepler high-multiplicity systems are drawn from the same intrinsic distribution. Clustering reveals a sub-population of systems with $\C\gtrsim0.33$ and wide spacings (large $\mathcal{S}$) that can be explained if a significant fraction of systems ($\sim20\%$) host additional undetected planets intermediate to the known planets. The distribution of $f$ is consistent with this hypothesis.
\end{itemize}

These findings highlight the success of our system-level approach for teasing out trends that cannot be seen by considering pairwise statistics alone. As we continue to refine this classification scheme, we anticipate the identification of finer-grained structure in the system-level parameter space. Such a detailed inter-system analysis will allow us to place these morphological descriptions in context with formation theories. To do so will require identifying new measures which highlight the most important physical characteristics of exoplanetary systems.

One important parameter has been conspicuously absent from our above discussion: eccentricity. A full set of system-level measures must include a term to capture the degree of eccentricity of a system. A candidate is the angular momentum deficit \citep[AMD;][]{Laskar1997, Laskar2017} , defined as the amount of angular momentum that would need to be added to a system in order to circularize all of the planets. At the moment, high-precision eccentricity measurements are few and far between \citep{Mills2019}. To remedy this deficiency, we are currently undertaking a project to refit \Kepler lightcurves for high-multiplicity systems that will yield improved estimates for inclinations and eccentricities. It may turn out that the best measure for describing exoplanet systems is not to use eccentricity and inclination independently, but rather to combine them into a sort of ``dynamical hotness'' term. Indeed, our $f$ value has some sensitivity to eccentricity as well as to inclination.

Many other measures of system-level architecture are possible. One possible candidate term is the typical distance from resonance for planets in a system. Such a resonance term could be expressed either as an average distance from resonance, or as a complexity-like term than answers the question ``if one planet in a system is near resonance, what is the likelihood that all other planets in that system are also near resonance?'' Other more exotic descriptors (each of which would require new measurements which are presently unobtainable) might be complexity terms based on alignments of nodes or periapses, varieties of bulk densities, atmospheric mass fractions, or even observed chemistry. \review{We also expect our existing quantities to evolve as more data become available.}

Future research should apply this work in several directions:
(1) using other populations of systems \review{(e.g. systems detected by RV or astrometry, or dynamically packed systems only)},
(2) computing our measures for population synthesis and planet formation models,
(3) obtaining better measurements of planetary inclinations,
(4) searching for missing planets in the systems which we have identified as likely to be hosting additional undetected planets; our hypothesis that the outlying high-complexity cluster is due to missing planets can also be tested using the RV planet sample, which is subject to different observational biases.
\review{(5) expanding our framework to include measures of eccentricity, planet bulk composition, and nearness to resonance}
\review{(6) investigating the sensitivity of inferences using different mass-radius relations.}

We look forward to putting the Solar System in a wider context - not only with regard to its planets system but also in relation to its giant moon systems. Our method provides a statistical target for planet formation models, no longer requiring the tuning of models to match just one system, e.g., the Solar System, TRAPPIST-1, or some other peculiar system of interest. Just as Galileo used the Jovian satellite system as a conceptual model for the Copernican Solar System, by looking at a much larger sample of exoplanetary systems, we can begin to see the system-level trends and whether such an identification has strong foundations.

\acknowledgements
We thank Leslie Rogers, Rebecca Willett, Fred Ciesla, Andrey Kravtsov, David Kipping, and Eric Ford for helpful discussions about our methodology. We thank an anonymous referee and the AJ statistics editor for helpful comments which improved the quality of this manuscript. We are especially grateful to Matthias He for guidance with using SysSim, to Gijs Mulders for providing synthetic EPOS catalogues, and to Andrew Neil for providing mass estimates for the planets; we thank all three for many insightful conversations which greatly enhanced the quality of this project.

We acknowledge support of grant NASA-NNX17AB93G through NASA’s Exoplanet Research Program.

Software: \texttt{scipy} \citep{Scipy}, \texttt{astropy} \citep{Astropy}, \texttt{scikit-learn} \citep{Scikit-learn}, \texttt{KDEpy}, \texttt{dcor}.

\bibliography{ms} \bibliographystyle{apj}

\end{document}